\documentclass[a4paper,11pt]{article}
\pdfoutput=1 

\usepackage{jheppub} 

\usepackage[T1]{fontenc} 

\usepackage{amssymb}

\title{\boldmath The simplest of them
  all: $t\bar{t} W^\pm$ at NLO accuracy in QCD}

\author[a]{Giuseppe Bevilacqua,}
\author[b]{Huan-Yu  Bi,}
\author[c]{Heribertus Bayu Hartanto,}
\author[d]{Manfred  Kraus}
\author[b]{and Malgorzata  Worek}

\affiliation[a]{MTA-DE Particle Physics Research Group, University of
  Debrecen, H-4010 Debrecen, \\PBox 105, Hungary}
\affiliation[b]{ Institute for Theoretical Particle Physics
and Cosmology, RWTH Aachen University, \\D-52056 Aachen, Germany}
\affiliation[c]{Institute for
Particle Physics Phenomenology, Department of Physics, 
Durham University, \\Durham, DH1 3LE, UK} 
\affiliation[d]{Physics Department, Florida State University,
Tallahassee, FL 32306-4350, USA}

\emailAdd{giuseppe.bevilacqua@science.unideb.hu}
\emailAdd{bihy@physik.rwth-aachen.de}
 \emailAdd{heribertus.b.hartanto@durham.ac.uk}
\emailAdd{mkraus@hep.fsu.edu} 
 \emailAdd{worek@physik.rwth-aachen.de}

 \abstract{
 Recent measurements of the $pp\to t\bar{t}W^\pm$ process in
multi-lepton final states, as performed by the ATLAS collaboration in
the context of the Higgs boson studies in the $t\bar{t}H$ channel,
have shown discrepancies between theoretical predictions and
experimental data. Such discrepancies have been observed both in the
overall normalisation as well as in the modelling of the
$t\bar{t}W^\pm$ process. With the goal of understanding and resolving
the modelling issues within the SM $t\bar{t}W^\pm$ process we report
on the state-of-the-art NLO QCD computation for this process.
Specifically, we calculate higher-order corrections to the $e^+ \nu_e
\,\mu^-\bar{\nu}_\mu \, e^+ \nu_e \, b\bar{b}$ and $e^- \bar{\nu}_e \,
\mu^+ {\nu}_\mu \, e^- \bar{\nu}_e \, b\bar{b}$ final state at the LHC
with $\sqrt{s}=13$ TeV. In the computation off-shell top quarks are
described by Breit-Wigner propagators, furthermore,
double-, single- as well as non-resonant top-quark contributions along
with all interference effects are consistently incorporated at the
matrix element level. Results at NLO QCD accuracy are presented in the
form of fiducial integrated and differential cross sections for two
selected renormalisation and factorisation scale choices and three
different PDF sets. The impact of the top quark off-shell effects on
the $t\bar{t}W^\pm$ cross section is also examined by an explicit
comparison to the narrow-width approximation.}

\dedicated{\rm TTK-20-13,  P3H-20-020, IPPP/20/14}


\begin{document} 
\maketitle
\flushbottom

%
\section{Introduction}
\label{sec:introduction}
%

Given the present values of collision energy and integrated luminosity
at the Large Hadron Collider (LHC), the observation of the associated
production of top quark pairs with a $W^\pm$ boson becomes
experimentally more and more accessible
\cite{Aaboud:2016xve,Sirunyan:2017uzs,
ATLAS:2019nvo,Aaboud:2019njj}. The immense amount of available phase
space leads to production and identification of all top quark final
states. Consequently, the LHC gives us finally the opportunity to
scrutinise not only the strength but also the structure and the
dynamics of $t\bar{t}W^\pm$ production.  The $t\bar{t}W^\pm$ process
allows for a direct measurement of the top quark coupling to $W^\pm$
bosons as well as the study of the top quark charge asymmetry
$(A^t_c)$ \cite{Maltoni:2014zpa}. At the leading order (LO) in
perturbation theory the $t\bar{t}W^\pm$ production process can only
occur via a $q\bar{q}^{\, \prime}$ annihilation, thus, contributions
from gluons in the initial states are not possible. The
$gq/g\bar{q}^{\, \prime}$ channels open up at next-to-leading order
(NLO) in QCD but the $gg$ production starts to be available only once
the next-to-next-to-leading order (NNLO) in QCD contributions are
incorporated. The absence of the symmetric $gg$ channel in the leading
terms of the perturbative expansion makes the resulting top quark
charge asymmetry as evaluated at the NLO level significantly larger
than in $t\bar{t}$ production. Thus, $t\bar{t}W^\pm$ can provide a
powerful complementary way to measure $A^t_c$.  The $t\bar{t}W^\pm$
process comprises multiple charged leptons, $b$-jets and missing
transverse momentum due to neutrinos. As a result, besides $A_c^t$
also the integrated charge asymmetry for the top decay products can be
examined at the LHC, namely the $b$-jet asymmetry $(A^b_c)$ and the
charged lepton asymmetry $(A^\ell_c)$. Both asymmetries are very large
and already present at the LO for this process. The polarisation and
asymmetry effects in $t\bar{t}W^\pm$ production can additionally offer
a useful handle to constrain new physics effects
\cite{Maltoni:2014zpa}.

Furthermore, the $t\bar{t}W^\pm$ process constitutes an important
background for the associated production of the Standard Model (SM)
Higgs boson and the top quark pair
\cite{Maltoni:2015ena,Sirunyan:2018hoz,Aaboud:2018urx}.  Analyses of
$t\bar{t}H$ and $t\bar{t}W^\pm$ production in multi-lepton final
states, which have been recently performed by the ATLAS collaboration,
have shown an overall higher normalisation for the $t\bar{t}W^\pm$
process \cite{ATLAS:2019nvo} when compared with theoretical
predictions provided by \textsc{OpenLoops} $+$ \textsc{Sherpa}
\cite{Gleisberg:2008ta, Cascioli:2011va} and/or
\textsc{MadGraph5${}_{-}$aMC$@$NLO} \cite{Alwall:2014hca}. The
normalisation factors obtained for the $t\bar{t}W^\pm$ background by
ATLAS for a very inclusive cut selection were $30\%-70\%$ higher than
the theoretical predictions. Additional problems have
been observed with the modelling of the final states in the phase
space regions, that are dominated by $t\bar{t}W^\pm$ production
\cite{ATLAS:2019nvo}. Specifically, shape disagreements between the
data and the $t\bar{t}W^\pm$ predictions from the Monte Carlo
simulations have been observed for various distributions. The most
important source of systematic uncertainties for the $t\bar{t}W^\pm$
process has been associated with the modelling of QCD radiation,
$b$-jet multiplicity and $W$ gauge boson charge asymmetry. The latter,
for example, has been studied with the help of the total charge
distribution.  Taking into account the impact of the assumptions made
on the $t\bar{t}W^\pm$ background modelling in the $t\bar{t}H$ cross
section measurement an improved description of the former is essential
to achieve higher precision in the future measurements.

Last but not least, the $t\bar{t}W^\pm$ process can lead to final
states that contain two charged leptons (electrons or muons) of the
same electric charge. Such signatures are referred to as same-sign
leptons and are a relatively rare phenomenon in the SM. 
The SM processes that can be a source of such final states are $W^\pm
Z$, $ZZ$, $W^\pm W^\pm jj$ as well as $t\bar{t}Z$ and $t\bar{t}W^\pm$
production. In the case of the $W^\pm W^\pm jj$ process two production
modes are possible: QCD-induced production and production via
vector-boson fusion. Finally, $WWW$, $W^+W^- Z$ and $ZZZ$ in various
decay channels give rise to same-sign leptons with either additional
leptons or jets. Even though SM processes leading to same-sign lepton
final states have usually very small cross sections, they are
indispensable in searches for physics beyond the SM (BSM). These BSM
searches are often focused on the presence of two same-sign leptons,
missing transverse momentum and two (light- or $b$-) jets
\cite{Aad:2016tuk,Khachatryan:2016kod,Sirunyan:2017uyt,
Aaboud:2017dmy}. The same-sign leptons signature is present in many
new physics scenarios, among others, in $R$-conserving SUSY models and
also in those with the explicit $R$-parity breaking
\cite{Barnett:1993ea,Guchait:1994zk,Baer:1995va,Dreiner:2006sv}. Similar
signatures are additionally predicted by non-SUSY models such as
minimal Universal Extra Dimensions \cite{Cheng:2002ab}. Same-sign
leptons are also the signature for top-quark partners that are
predicted in models where the Higgs is a pseudo-Goldstone boson
\cite{Contino:2008hi}. They are essential for the production of the
doubly charged Higgs bosons in the left-right symmetric model and in
the Higgs triplet model \cite{Maalampi:2002vx}. Furthermore, same-sign
leptons are unavoidable when searching for heavy Majorana neutrinos
\cite{Almeida:1997em} and same-sign top quark pair resonances
\cite{Bai:2008sk}.

 Given such rich phenomenological applications it is essential to
describe all features of the $t\bar{t}W^\pm$ process as accurately as
possible, in order to either deepen our understanding of the SM or
maximise the sensitivity to deviations from it. To achieve this,
theoretical calculations for the $t\bar{t}W^\pm$ process need to
comprise all quantum effects already at the matrix element level. In
addition, the size of higher-order corrections and theoretical
uncertainties have to be carefully scrutinised in such a complex
environment.  Moreover, the choice of  appropriate renormalisation and
factorisation scales must be addressed to keep the process under
excellent theoretical control.

The goal of this paper is, therefore, to provide  state-of-the-art
NLO QCD predictions for the SM $t\bar{t}W^\pm$ process in the
multi-lepton channel. More precisely, we shall calculate NLO QCD
corrections to the $e^+ \nu_e \,\mu^- \bar{\nu}_\mu \, e^+
\nu_e \, b\bar{b}$ and $e^- \bar{\nu}_e \, \mu^+ {\nu}_\mu \, e^-
\bar{\nu}_e\, b\bar{b}$ final state. In the calculation,  for the first
time,  all double-, single- and non-resonant Feynman diagrams will be
consistently taken into account together with the off-shell effects of
the top quarks. Additionally, non-resonant and off-shell effects
related  to the $W$ gauge bosons will be incorporated. This
calculation constitutes the first fully realistic NLO QCD computation
for top quark pair production with the additional $W^\pm$ gauge boson.

As a final comment, we note that NLO QCD corrections to the inclusive
$t\bar{t}W^\pm$ process, with the on-shell top quarks, have been
calculated for the first time in \cite{Hirschi:2011pa} and afterwards
recomputed in \cite{Maltoni:2014zpa,Maltoni:2015ena}. Theoretical
predictions for $t\bar{t}W^\pm$ at NLO in QCD have been additionally
matched with shower MC programs using either the \textsc{Powheg}
method or the MC@NLO framework
\cite{Garzelli:2012bn,Maltoni:2015ena}. In all cases top quark and $W$
gauge boson decays have been treated in the parton shower
approximation omitting the NLO $t\bar{t}$ spin correlations. Finally,
in Ref. \cite{Campbell:2012dh} improved calculations for the process
have been presented. Specifically, NLO QCD corrections to the
production and decays of top quarks and $W$ gauge bosons have been
included with full spin correlations in the narrow-width approximation
(NWA). Besides NLO QCD corrections, a further step towards a more
precise modelling of the on-shell $t\bar{t}W^\pm$ production process
has been achieved by including either NLO electroweak corrections
\cite{Frixione:2015zaa} and the subleading electroweak corrections
\cite{Dror:2015nkp,Frederix:2017wme} or by incorporating soft gluon
resummation effects with the next-to-next-to-leading logarithmic
(NNLL) accuracy
\cite{Li:2014ula,Broggio:2016zgg,Kulesza:2018tqz}. Very recently
subleading electroweak corrections together with the $t\bar{t}$
spin-correlation effects for the on-shell $t\bar{t}W^\pm$ production
matched to parton shower programs have been examined in the
multi-lepton channel \cite{Frederix:2020jzp}. Top quark and $W$ gauge
boson decays have been realised via parton showers within the
\textsc{MadSpin} framework \cite{Artoisenet:2012st}, which allowed to
account for the leading-order spin correlations.

The paper is organised as follows. In Section \ref{sec:methods} we
briefly outline the framework of the calculation and discuss
cross-checks that have been performed.  The theoretical setup for LO
and NLO QCD results is given in Section
\ref{sec:setup}. Phenomenological results for $t\bar{t}W^+$ are
discussed in detail in Section \ref{sec:ttwp}. They are provided for
the LHC centre of mass system energy of $\sqrt{s}=13$ TeV and for two
renormalisation and factorisation scale choices as well as for the
following three parton distribution functions (PDFs): NNPDF3.0, MMHT14
and CT14. Theoretical uncertainties due to the scale dependence and
PDFs are also discussed in Section \ref{sec:ttwp} both for the
integrated and differential fiducial $t\bar{t}W^+$ cross
sections. Additionally, in Section \ref{sec:ttwp} the impact of the
off-shell effects on the $t\bar{t}W^+$ cross section is
examined. Section \ref{sec:ttwm} is devoted to results for
$t\bar{t}W^-$ production.  In this case only theoretical predictions
for the integrated fiducial cross sections are presented so as not to
extend the length of the manuscript unnecessarily. This is well
justified by the fact that the NLO QCD effects for $t\bar{t}W^+$ and
$t\bar{t}W^-$ are very similar. Finally, in Section \ref{sec:sum} our
results for the $t\bar{t}W^\pm$ production process are summarised and
conclusions are outlined.

%
\section{Outline of the calculations and cross-checks}
\label{sec:methods}
%

%
\begin{figure}[t!]
\begin{center}
  \includegraphics[width=1.0\textwidth]{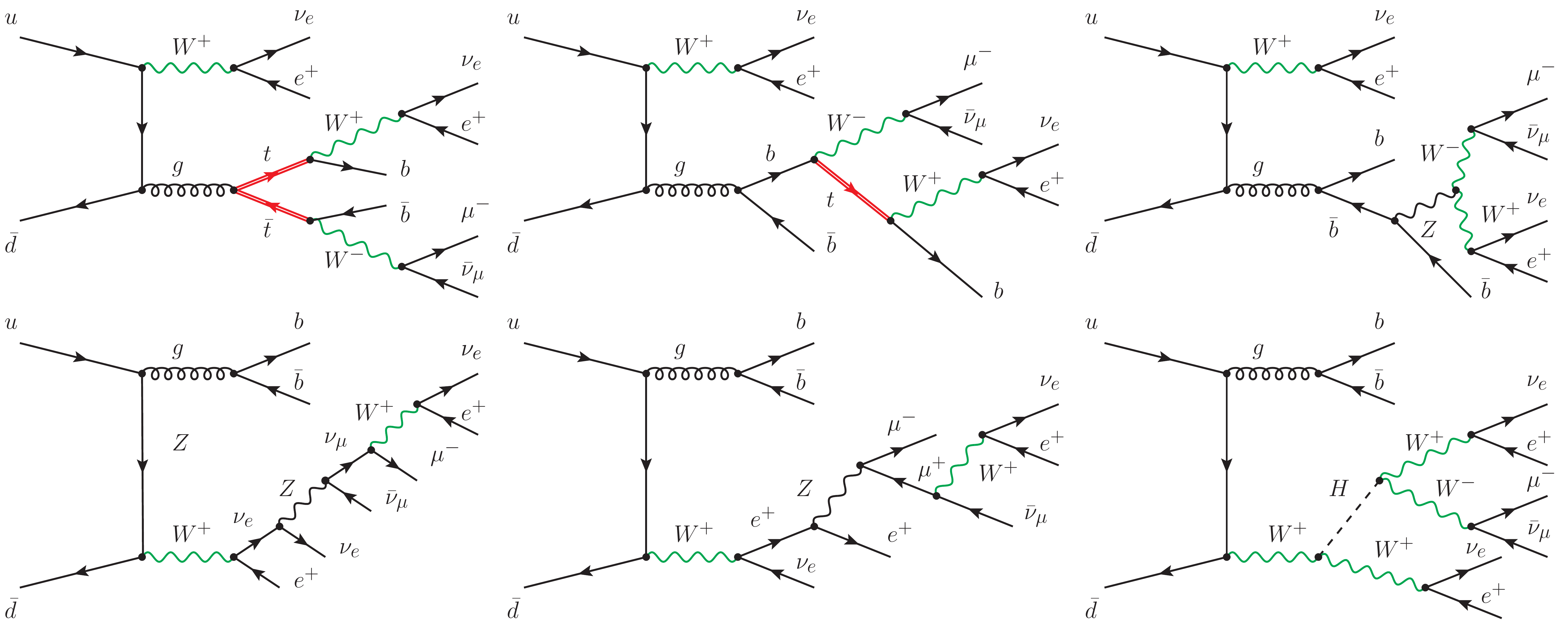}
\end{center}
\vspace{-0.4cm}
\caption{\label{fig:fd} \it
  Representative tree level Feynman diagrams for the $pp\to e^+\nu_e\,
\mu^-\bar{\nu}_\mu\, e^+\nu_e \, b\bar{b}$ process at ${\cal
O}(\alpha_s^2\alpha^6)$. In the first row diagrams with two (first
diagram), only one (second diagram), or no (last diagram) top-quark
resonances are presented.  The double lines indicate the top and
anti-top quark.  In the second row diagrams that involve the $W$ gauge
boson resonance are given (first and the second diagram). They
contribute to the finite $W$ width corrections. The last diagram in
the second row comprises the Higgs-exchange contribution that appears
even though the $b$ quarks are treated as massless partons.}
\end{figure}
%

We compute the NLO QCD corrections to the full hadronic
process $pp \to e^+\nu_e\, \mu^-\bar{\nu}_\mu\, e^+\nu_e \,
b\bar{b}$. We consider the tree-level amplitude at ${\cal
O}(\alpha_s^2\alpha^6)$. The $e^+\nu_e\, \mu^-\bar{\nu}_\mu\, e^+\nu_e
\, b\bar{b}$ final state\footnote{We shall concentrate here on the
$t\bar{t}W^+$ process, however, a similar description applies to
$t\bar{t}W^-$ production. We note here that the integrated fiducial
cross section for $t\bar{t}W^+$ is larger than the one for
$t\bar{t}W^-$. This can be easily understood by looking at the 
partonic subprocesses  and the corresponding PDFs as well as their impact
on the $pp$ collisions at the LHC with $\sqrt{s}=13$ TeV. We also note
that at LO $t\bar{t}W^-$ is produced via $\bar{u}d$ and
$\bar{c}s$.} is produced via the scattering of one up-type quark and
the corresponding down-type anti-quark.  The quark-gluon initial
state opens up only at the next order in $\alpha_s$. Due to the large
gluon luminosity this might have a potentially large impact on the
size of the higher-order corrections and theoretical uncertainties.
Unlike for the processes $pp \to t\bar{t}j /t\bar{t}\gamma/
t\bar{t}Z/t\bar{t}H$, the production process $t\bar{t}W^+$ only
originates from the gluon-gluon initial state starting from
NNLO. At the LO, however, we distinguish 4 partonic
subprocesses for the full hadronic process. All partonic subprocesses
can be obtained from $q\bar{q}^{\,\prime} \to e^+\nu_e\,
\mu^-\bar{\nu}_\mu\, e^+\nu_e \, b\bar{b}$  by substituting different
quark flavours $(q=u,d,c,s)$. Each subprocess involves $556$ tree
Feynman diagrams.  Examples of Feynman diagrams for the $u\bar{d} \to
e^+ \nu_e \mu^- \bar{\nu}_\mu e^+ \nu_e b\bar{b}$ partonic subprocess
are depicted in Figure \ref{fig:fd}.  Even though we treat $b$ quarks
as massless partons there are Higgs-boson-exchange Feynman diagrams,
see e.g. the last diagram in the second row of Figure
\ref{fig:fd}. Once this contribution is also taken into account the
number of diagrams increases to $564$. To regularise intermediate
top-quark resonances in a gauge-invariant way we employ the
complex-mass scheme \cite{Denner:1999gp,
Denner:2005fg,Bevilacqua:2010qb,Denner:2012yc}, which consistently
describes off-shell top quark contributions by the Breit-Wigner
distribution. All matrix elements are evaluated using the complex
top-quark mass $\mu_t$ defined by
\begin{equation}
\mu_t^2=m_t^2-im_t\Gamma_t\,.
\end{equation}
The $W$ and $Z$ gauge bosons, on the other hand, are
treated within the fixed width scheme, see e.g. \cite{Kauer:2001sp}.
Since we are interested in NLO QCD corrections the naive use of
Breit-Wigner propagators for gauge bosons does not introduce problems
in the calculations. The preservation of gauge symmetries (Ward
Identities) by this approach has been explicitly checked up to the
one-loop level.  The calculation of the scattering amplitudes for the
$q\bar{q}^{\,\prime} \to e^+\nu_e\, \mu^-\bar{\nu}_\mu\, e^+\nu_e \,
b\bar{b}$ process is based on the well-known off-shell Dyson-Schwinger
iterative algorithm that is implemented within the \textsc{Helac-NLO}
framework \cite{Bevilacqua:2011xh} and in the \textsc{Helac-Phegas} MC
program \cite{Cafarella:2007pc}. The latter MC library is used to
cross check all LO results. Phase space integration is performed and
optimised with the help of \textsc{Parni} \cite{vanHameren:2007pt} and
\textsc{Kaleu} \cite{vanHameren:2010gg}.
%
\begin{table}[t!]
\begin{center}
\begin{tabular}{ccc}
\hline \hline 
  {One-loop correction type}& \quad \quad \quad
  &{Number of Feynman diagrams} \\[0.2cm]
  \hline \hline
 {Self-energy} && 7708 \\[0.2cm]
 {Vertex} &  &4236 \\[0.2cm]
 {Box-type}                && 2606   \\[0.2cm]
 {Pentagon-type} && 1116  \\[0.2cm]
  {Hexagon-type}              && 260 \\[0.2cm]
{Heptagon-type}  & &16    \\ [0.2cm]
  \hline \hline
{Total number}  && 15942 \\[0.2cm]
   \hline \hline
 \end{tabular}
\end{center}
\caption{\label{tab:one-loop} \it
  The number of one-loop Feynman diagrams for the $u\bar{d}\to
e^+\nu_e \mu^- \bar{\nu}_\mu e^+\nu_e \, b\bar{b}$ partonic
subprocess at ${\cal O}(\alpha_s^3 \alpha^6)$ for the $pp\to e^+ \nu_e
\mu^- \bar{\nu}_\mu  e^+ \nu_e \,  b\bar{b} +X$ process. The
Higgs boson exchange contributions are not taken into account  and the
Cabibbo-Kobayashi-Maskawa mixing matrix is assumed to be diagonal.}
\end{table}

The virtual corrections can be classified into self-energy, vertex,
box-type, pentagon-type, hexagon-type and heptagon-type
corrections. In Table \ref{tab:one-loop} we provide the number of
one-loop Feynman diagrams, that corresponds to each topology for the
following partonic subprocess $u \bar{d}\to e^+\nu_e\,
\mu^-\bar{\nu}_\mu\, e^+\nu_e \, b\bar{b}$. These numbers have been
generated with the help of the \textsc{Qgraf} program
\cite{Nogueira:1991ex}, which generates Feynman diagrams for various
types of QFT models. The 1-loop corrections have been evaluated by the
\textsc{Helac-1Loop} \cite{vanHameren:2009dr} MC library, which
incorporates \textsc{CutTools} \cite{Ossola:2006us,Ossola:2007ax} and
\textsc{OneLOop} \cite{vanHameren:2010cp}. We have cross-checked our
results with the publicly available general purpose MC program
\textsc{MadGraph5-aMC@NLO}. Specifically, we have compared results for
the virtual NLO contribution to the squared amplitude, $2\Re({\cal
M}^{*}_{\rm tree}{\cal M}_{\rm one-loop})$, for a few phase-space
points for the $u\bar{d}$ partonic subprocess. For all phase-space
points that we have tested perfect agreement has been found.

For the calculation of the real emission contributions, the package
\textsc{Helac-Dipoles} \cite{Czakon:2009ss} is employed. It comprises
the dipole formalism of Catani and Seymour
\cite{Catani:1996vz,Catani:2002hc} for arbitrary helicity eigenstates
and colour configurations of the external partons and the Nagy- Soper
subtraction scheme \cite{Bevilacqua:2013iha}, which makes use of
random polarisation and colour sampling of the external partons. Two
independent subtraction schemes allow us to cross check the
correctness of the real corrections by comparing the two
results. Furthermore, a restriction on the phase space of the
subtraction term is considered for both Catani-Seymour and Nagy-Soper
schemes and additionally used for cross checks, see
Ref.~\cite{Bevilacqua:2009zn} and Ref.~\cite{Czakon:2015cla} for
technical details of our implementation.  The real
correction process $pp \to e^+\nu_e\, \mu^-\bar{\nu}_\mu\, e^+\nu_e \,
b\bar{b}j$ receives contributions from the following 12 partonic
subprocesses
\begin{equation}
  \begin{split}
    &gq  \,\,  \to   e^+\nu_e\,
      \mu^-\bar{\nu}_\mu\, e^+\nu_e \, b\bar{b} \,q^{\, \prime} \,,
      \\[0.2cm]
      &g \bar{q}^{\, \prime}  \to  e^+\nu_e\,
      \mu^-\bar{\nu}_\mu\, e^+\nu_e \, b\bar{b} \,\bar{q}\,, \\[0.2cm]
      &q \bar{q}^{\,\prime} \to   e^+\nu_e\,
      \mu^-\bar{\nu}_\mu\, e^+\nu_e \, b\bar{b} \,g \,.
     \end{split}
   \end{equation}
Each subprocess comprises 3736 Feynman diagrams. As for the number of
Catani-Seymour dipoles and Nagy-Soper subtraction terms we have
respectively $12$ and $4$ for the first two subprocesses as well as
$15$ and $5$ for the third  one.  The difference between the number of
Catani-Seymour dipoles and Nagy-Soper subtraction terms corresponds to
the total number of possible spectators that are only relevant in the
Catani-Seymour subtraction scheme.

To summarise, our computational system is based on
\textsc{Helac-1Loop} and \textsc{Helac- Dipoles}, which are both parts
of the \textsc{Helac-NLO} MC program. Let us note here, that among the
processes of associated $t\bar{t}$ production which have been
calculated so far with \textsc{Helac-NLO} ($t\bar{t}X$, where
$X=j,\gamma,Z,W^\pm$ \cite{Bevilacqua:2015qha,Bevilacqua:2016jfk,
Bevilacqua:2018woc, Bevilacqua:2019cvp}), $t\bar{t}W^\pm$ is perhaps
the simplest one in terms of computational complexity. We draw this
conclusion after comparing various criteria, such as the number of
Feynman diagrams and subtraction terms involved in the calculation, or
the number of partonic subprocesses and color structures of the
amplitudes. Yet, computing NLO QCD corrections for the $t\bar{t}W^\pm$
process with the complete off-shell effects included, is challenging
and requires a good computer cluster in order to accomplish the task
in a reasonable amount of time. For this reason we store our
theoretical predictions in the form of events, available in the format
of either (modified) Les Houches Event Files \cite{Alwall:2006yp} or
ROOT Ntuples \cite{Antcheva:2009zz}. Expanding on methods
presented in Ref.~\cite{Bern:2013zja}, each event is stored with
additional matrix-element and PDF information which allows on-the-fly
reweighting for different choices of scales and PDFs. In this way one
can obtain predictions for arbitrary infrared-safe observables,
kinematical cuts, renormalisation/factorisation scales and PDFs,
without requiring additional rerunning of the computationally  intense
\textsc{Helac-Nlo} code. A user-friendly program, named
\textsc{HEPlot} \cite{heplot}, has been developed to easily obtain physical
predictions out of these event files. Both the event files and the
\textsc{HEPlot} program are available upon request and might be
directly used for experimental analyses at the LHC as well as to
obtain accurate SM predictions in phenomenological studies on, e.g.,
Higgs boson or BSM physics.

%
\section{LHC setup}
\label{sec:setup}
%

We start with the $t\bar{t}W^+$ production process that is calculated
at NLO in QCD for the LHC Run II energy of $\sqrt{s}=13$ TeV.
Specifically, the following final state is considered: $e^+\nu_e\,
\mu^-\bar{\nu}_\mu\, e^+\nu_e \, b\bar{b}+X$ at perturbative order
${\cal O}(\alpha_s^3\alpha^6)$. By choosing different lepton
generations for $W^+ \to e^+ \nu_e$ and $ W^-  \to \mu^-
\bar{\nu}_\mu$ we avoid virtual photon singularities
stemming from the $\gamma \to e^+e^-$ and $\gamma \to \mu^+ \mu^-$
decays. However, we have checked by an explicit LO calculation that
these interference effects are at per-mil level. The complete cross
section for the $pp \to \ell^+ \nu_\ell \, \ell^- \bar{\nu}_\ell \,
\ell^+ \nu_\ell \, b\bar{b}$ process, where $\ell^\pm$ stands for
$\ell^\pm= e^\pm,\mu^\pm$, can be obtained by multiplying the results
from this paper with a lepton-flavour factor of $8$. We do not take into
account the $\tau$ leptons. The large variety of final states into
which the tau leptons can decay makes them very challenging to
reconstruct and identify at hadron colliders
\cite{ATLAS:2017mpa,Khachatryan:2015dfa}. For this reason they are
often studied  separately at the LHC. Additionally, we have examined
the impact of the Higgs boson contributions on the $pp\to e^+\nu_e\,
\mu^-\bar{\nu}_\mu\, e^+\nu_e \, b\bar{b}$ fiducial cross section.  We
have checked that, at LO with $m_H=125$ GeV and $\Gamma_H= 4.07 \times
10^{-3}$ GeV, the latter contribute at the level of per-mille.  Furthermore, for a
variety of differential distributions, which we have examined,
differences between theoretical results with and without these
contributions were within the integration errors for our
setup. Consequently, in the following we shall neglect the Higgs boson
contribution both at the LO and NLO. For our calculation we keep the
Cabibbo-Kobayashi-Maskawa (CKM) mixing matrix diagonal. We have
checked, however, the impact of off-diagonal contributions on the
fiducial cross sections using LO and NLO calculations in the NWA.  We
use the approximation for the CKM matrix that considers mixing only
between the first two generations of quarks,  with the Cabibbo angle
$\sin\theta_C=0.225686$. By employing \textsc{Helac-NLO} and taking
into consideration additional subprocesses we have established that
off-diagonal contributions are at the $2\%$ level at LO and below
$1.5\%$ at NLO. These findings have been cross-checked with the
\textsc{Mcfm} Monte Carlo program \cite{Campbell:2019dru}. Following
recommendations of the PDF4LHC Working Group for the usage of PDFs
suitable for applications at the LHC Run II \cite{Butterworth:2015oua}
we employ CT14 \cite{Dulat:2015mca},
MMHT14 \cite{Harland-Lang:2014zoa} and NNPDF3.0 \cite{Ball:2014uwa}.
In particular, we use \texttt{NNPDF30-nlo-as-0118} with
$\alpha_s(m_Z)=0.118$ (\texttt{NNPDF30-lo-as-0130} with
$\alpha_s(m_Z)=0.130$) as the default PDF set at NLO (LO). In
addition, we present results for \texttt{CT14nlo} and
\texttt{MMHT14nlo68clas118} at NLO as well as \texttt{CT14llo} and
\texttt{MMHT14lo68cl} at LO. The running of the strong coupling
constant $\alpha_s$ with two-loop (one-loop) accuracy at NLO (LO) is
provided by the LHAPDF interface \cite{Buckley:2014ana}. The number of
active flavours is set to $N_F = 5$ and the following SM parameters
are used 
\begin{equation}
\begin{array}{lll}
 G_{ \mu}=1.166378 \cdot 10^{-5} ~{\rm GeV}^{-2}\,, &\quad \quad \quad
&   m_{t}=172.5 ~{\rm GeV} \,,
\vspace{0.2cm}\\
 m_{W}=80.385 ~{\rm GeV} \,, &
&\Gamma^{\rm NLO}_{W} = 2.09767 ~{\rm GeV}\,, 
\vspace{0.2cm}\\
  m_{Z}=91.1876  ~{\rm GeV} \,, &
&\Gamma^{\rm NLO}_{Z} = 2.50775 ~{\rm GeV}\,.
\end{array}
\end{equation}
For the $W$ and $Z$ gauge boson width, $\Gamma^{\rm NLO}_{W}$ and
$\Gamma^{\rm NLO}_{Z}$,  we use the NLO
QCD values as calculated respectively for $\mu_R=m_W$ and
$\mu_R=m_Z$. We utilise them for LO and NLO matrix elements. All other
partons, including bottom quarks, and leptons are treated as massless
particles. The LO and NLO top quark widths  for the off-shell case are
calculated according to formulae from 
Ref. \cite{Jezabek:1988iv,Chetyrkin:1999ju,Denner:2012yc} and are
given by
\begin{equation}
\begin{array}{lll}
  \Gamma_{t, {\rm off-shell}}^{\rm LO} = 1.45759 ~{\rm GeV}\,, &\quad
                                                                 \quad
                                                                 \quad
  \quad&
  \Gamma_{t, {\rm off-shell}}^{\rm NLO} = 1.33247  ~{\rm GeV}\,.
\end{array}
\end{equation}
On the order hand, for the NWA case we use the following values 
\begin{equation}
\begin{array}{lll}                                                                                  
  \Gamma^{\rm LO}_{t,{\rm NWA}} = 1.48063 ~{\rm GeV}\,, &\quad \quad
                                                          \quad
                                                          \quad
  & \Gamma^{\rm
NLO}_{t,{\rm NWA}} = 1.35355 ~{\rm GeV}\,.
\end{array}
\end{equation}
The top quark width is treated as a fixed parameter throughout this
work. Its value corresponds to a fixed scale $\mu_R=m_t$, that
characterises the top quark decays, and is equal to
$\alpha_s(m_t)=0.107671$. The $\alpha_s(m_t)$ parameter is independent
of $\alpha_s(\mu_0)$ that goes into the matrix element calculations as
well as PDFs, since the latter describes the dynamics of the whole
process. Let us add here as well that, while calculating the scale
dependence for the NLO cross section, $\Gamma_t^{\rm NLO}$ is kept
fixed independently of the scale choice.  The error
introduced by this treatment is, however, of higher order and
particularly for two scales $\mu = m_t/2$ and $\mu = 2m_t$ is below
$1.5\%$ as we have checked by the explicit NLO calculation in the
NWA. Consequently, omitting the variation of $\Gamma^{\rm NLO}_{t,{\rm
NWA}}$ can underestimate the NLO scale dependence maximally by
$1.5\%$. The electromagnetic coupling $\alpha$ is calculated from the
Fermi constant $G_\mu$, i.e.  in the $G_\mu-$scheme, via
\begin{equation}
\alpha_{G_\mu}=\frac{\sqrt{2}}{\pi} \,G_F \,m_W^2  \,\sin^2\theta_W
\,,
\end{equation}
where $\sin^2\theta$ is defined according to  
\begin{equation}
\sin^2\theta = 1-\frac{m_W^2}{m_Z^2}\,,
\end{equation}
and $G_\mu$ is extracted from the muon decay. Fixed-order calculations
at NLO in QCD contain a residual dependence on the renormalisation
$(\mu_R)$ and the factorisation scale $(\mu_F)$. This dependence
arises from the truncation of the perturbative expansion in
$\alpha_s$. For that reason observables depend on the values of
$\mu_R$ and $\mu_F$. They have to be provided as input
parameters, and can generally be functions of the
external momenta.  The uncertainty on higher orders is estimated by
varying $\mu_R$ and $\mu_F$ independently around a central scale
$\mu_0$ in the range
\begin{equation}
  \label{scale}
  \frac{1}{2}  \le \frac{\mu_R}{\mu_0}\,, \frac{\mu_F}{\mu_0} \le  2 \,.
\end{equation}
It is conventional to require  the following additional condition to be met
\begin{equation}
\frac{1}{2}\le
\frac{\mu_R}{\mu_F} \le  2 \,.
\end{equation}
We search for the minimum and maximum of the resulting cross
sections.  Because none of the ratios $\mu_F/\mu_0$, $\mu_R/\mu_0$ and
$\mu_R/\mu_F$ can be larger than two or smaller than one-half it is
sufficient  to consider the following pairs only
\begin{equation}
\label{scan}
\left(\frac{\mu_R}{\mu_0}\,,\frac{\mu_F}{\mu_0}\right) = \Big\{
\left(2,1\right),\left(0.5,1  
\right),\left(1,2\right), (1,1), (1,0.5), (2,2),(0.5,0.5)
\Big\} \,.
\end{equation}
For the central value of $\mu_0$ we consider two cases. First, we
employ a fixed scale given by
\begin{equation} \mu_0= m_t+\frac{m_W}{2} \,.
 \end{equation}
The scale choice $\mu_0=m_t+m_V/2$, where $V$ stands for a massive
boson $(V=H,Z,W^\pm)$, has previously been used in higher order
calculations for $pp\to t\bar{t}V$ production with on-shell top quarks
\cite{Campbell:2012dh,Beenakker:2001rj,Beenakker:2002nc,Dawson:2002tg,
Dawson:2003zu,Lazopoulos:2008de,Kardos:2011na}. Thus, we follow this
prescription as well. Our second choice for the scale is dynamical,
i.e. phase-space dependent. The scale is chosen to be the
scalar sum of all transverse momenta in the event, including the
missing transverse momentum. We denote this scale as $H_T$. Not only
the functional form of $\mu_0$ is important but also the overall
factor that stands in front. To this end, we select
\begin{equation}
  \mu_0=\frac{H_T}{3} \,,
\end{equation}
where $H_T$ is given by
\begin{equation}
  \label{ht}
H_T= p_T(\ell_1 )+ p_T(\ell_2)+ p_T(\mu^-) +p_{T}^{miss}
+ p_{T} (j_{b_1}) + p_{T} (j_{b_2}) \,,
\end{equation}
where $\ell$ labels positrons.  The choice we make is blind to the
fact that in the $pp \to e^+\nu_e \mu^-\bar{\nu}_\mu e^+\nu_e \,
b\bar{b}$ process top-quark resonances might appear. Thus, it seems to
be a more natural option for the process with the complete top-quark
off-shell effects included. It should play a vital role
especially in the case of various dimensionful observables in the high
$p_T$ phase space regions where the single- and non-resonant
contributions comprise a significant part of the integrated cross
section. At times they can even be larger than the double-resonant
contribution.  We note that $H_T$ from Eq.~\eqref{ht} is directly
measurable, i.e. it is defined with the help of observable final
states that pass all the cuts that we shall specify in the following.
Furthermore, since the electron and the muon reconstruction and charge
identification can be performed at the LHC with very high efficiency
\cite{Aad:2016jkr,Aaboud:2019ynx}, we can distinguish between $\mu^-$
and $e^+$ in our studies. To differentiate between the two positrons,
however, the ordering in $p_T$ is introduced. The same applies to the
two $b$-jets that are present in the final state. Consequently, in
Eq.~\eqref{ht} $j_{b_1}$ and $j_{b_2}$ stand for the hardest and the
softest $b$-jet, $\mu^-$ labels the muon, $\ell_{1,2}$ corresponds to
the hardest and the softest positron and $p_{T}^{miss}$ is the missing
transverse momentum, which is built out of two $\nu_e$'s and a
$\bar{\nu}_\mu$.  We define jets out of all final-state partons with
pseudo-rapidity $|\eta| <5$. In particular, partons are recombined
into jets via the IR-safe {\it anti}$-k_T$ jet algorithm
\cite{Cacciari:2008gp} where the separation parameter $R=0.4$ is used.
We require exactly two $b$-jets and three charged leptons, two of
which are same-sign charged leptons. All final states have to fulfil
the following selection criteria that mimic the ATLAS detector
response
\begin{equation}
\begin{array}{ll l l  }
  p_{T} ({\ell})>25 ~{\rm GeV}\,,    &
                                     \quad\quad \quad\quad\quad &
                                                                  p_{T} (j_b)>25 ~{\rm GeV}\,, 
\vspace{0.2cm}\\
 |y(\ell)|<2.5\,,&& |y(j_b)|<2.5 \,,   
\vspace{0.2cm}\\
 \Delta R(\ell  \ell) > 0.4\,, &&
\Delta R (\ell \, j_b) > 0.4\,, 
\end{array}
\end{equation}
where $\ell$ stands for the charged lepton $\ell =\mu^-,e^+$. Such
selection would ensure well observed isolated charged leptons  and
$b$-jets in the central rapidity regions of the ATLAS detector. We put no
restriction on the kinematics of the extra (light) jet and the missing
transverse momentum.

%
\section{\boldmath Phenomenological results for
  $t\bar{t}W^+$}
\label{sec:ttwp}
%

%
\subsection{Fiducial  cross sections}
%

We generate theoretical predictions for the LHC that is a $pp$
collider, thus, the rates for $t\bar{t}W^+$ and $t\bar{t}W^-$ are not
equal.  We start with the $t\bar{t}W^+$ production process as it has
the largest cross section between the two. We begin the presentation
of our results with a discussion of the integrated fiducial cross
section for the fixed scale choice. With the input parameters and cuts
specified as in Section \ref{sec:setup}, we arrive at the following
predictions if the NNPDF3.0 PDF sets are employed
\begin{equation}
  \begin{split}
&\sigma^{\rm LO}_{e^+\nu_e\,
\mu^-\bar{\nu}_\mu\, e^+\nu_e \, b\bar{b}} \left({\rm NNPDF3.0},
\mu_0=m_t+m_W/2\right)=
106.9^{\,+27.7~(26\%)}_{\,-20.5~(19\%)} \, {\rm [scale]} \, {\rm ab}\,,
\\[0.2cm]
&\sigma^{\rm NLO}_{e^+\nu_e\, 
\mu^-\bar{\nu}_\mu\, e^+\nu_e \, b\bar{b}} \left({\rm NNPDF3.0},
\mu_0=m_t+m_W/2 \right)= 123.2^{\,+6.3~(5\%)}_{\, -8.7~(7\%)} \, {\rm
[scale]} \, {}^{+2.1~(2\%)}_{-2.1~(2\%)} \, {\rm [PDF]} \, {\rm ab}\,.
\end{split}
\end{equation}  
For the MMHT14 PDF sets we obtain instead
\begin{equation}
  \begin{split}
&\sigma^{\rm LO}_{e^+\nu_e\,
\mu^-\bar{\nu}_\mu\, e^+\nu_e \, b\bar{b}} \left({\rm MMHT14},
\mu_0=m_t+m_W/2\right)= 102.2^{\,+27.0~(26\%)}_{\,-19.9~(19\%)}
\, {\rm [scale]} \, {\rm ab}\,,
\\[0.2cm]
&\sigma^{\rm NLO}_{e^+\nu_e\, 
\mu^-\bar{\nu}_\mu\, e^+\nu_e \, b\bar{b}} \left({\rm MMHT14},
\mu_0=m_t+m_W/2 \right)= 123.1^{\, +5.9~(5\%)}_{\,-8.4~(7\%)}
\, {\rm
[scale]} \, {}^{+2.8~(2\%)}_{-2.5~(2\%)}\, {\rm [PDF]} \, {\rm ab}\,.
\end{split}
\end{equation}  
Finally, with the CT14 PDF sets our results are as follows
\begin{equation}
  \begin{split}
&\sigma^{\rm LO}_{e^+\nu_e\,
\mu^-\bar{\nu}_\mu\, e^+\nu_e \, b\bar{b}} \left({\rm CT14},
\mu_0=m_t+m_W/2\right)= 103.8^{\,+26.7~(26\%)}_{\,-19.7~(19\%)}
\, {\rm [scale]} \, {\rm ab}\,,
\\[0.2cm]
&\sigma^{\rm NLO}_{e^+\nu_e\, 
\mu^-\bar{\nu}_\mu\, e^+\nu_e \, b\bar{b}} \left({\rm CT14},
\mu_0=m_t+m_W/2 \right)= 122.9^{\, +6.0~(5\%)}_{\,-8.6~(7\%)}\, {\rm
[scale]} \, {}^{+3.0~(2\%)}_{-3.5~(3\%)} \, {\rm [PDF]} \, {\rm ab}\,.
\end{split}
\end{equation}  
We do not provide the LO PDF uncertainties because they are similar to
the NLO values, i.e. an order of magnitude smaller than the LO
theoretical uncertainties due to scale dependence. For the NNPDF3.0
PDF sets we obtain positive and moderate NLO QCD corrections of the
order of $15\%$.  For the MMHT14 PDF set instead we receive $20\%$ and
for CT14 $18\%$ corrections. Scale uncertainties taken as the maximum
of the lower and upper bounds are at the $26\%$ level at the LO. After
inclusion of the NLO QCD corrections, they are reduced down to
$7\%$. Another source of theoretical uncertainties comes from the PDF
parametrisation. Using the error PDF sets the NLO PDF uncertainties
have been calculated separately for NNPDF3.0, MMHT14 and CT14. They
are rather small at the level of $2\%-3\%$.  We should mention here,
that the CT14 PDF uncertainties are provided as $90\%$ confidence
level intervals, therefore, we have rescaled them by a factor $1.645$
to compare with other PDF sets, for which uncertainties are provided
as $68\%$ confidence level intervals.  We can further notice, that NLO
results for three different PDF sets are very consistent as the
differences among them are at the per-mill level only. Overall, the
PDF uncertainties for the process under consideration are well below
the theoretical uncertainties due to the scale dependence, which
remain the dominant source of the theoretical systematics.
%
\begin{table}[t!]
\begin{center}
  \begin{tabular}{cccccccc}
     \hline \hline
    &&&&&&&\\ 
    \multicolumn{8}{c}{$\mu_R=\mu_F=\mu_0=m_t+m_W/2$} \\
    &&&&&&&\\ 
    \hline \hline &&&&&&&\\ PDF &$p_{T}(j_b)$ & $\sigma^{\rm LO}$ [ab]
& $\delta_{\rm scale}$ & $\sigma^{\rm NLO}$ [ab] & $\delta_{\rm scale}$ &
$\delta_{\rm PDF}$ & $\sigma^{\rm NLO}/\sigma^{\rm LO}$ \\ &&&&&&&\\
\hline \hline &&&&&&&\\ CT14 & $25$ & $103.8$ &
$^{+26.7~(26\%)}_{-19.7~(19\%)}$ & $122.9$ &
$^{+6.0~(5\%)}_{-8.6~(7\%)}$ & $^{+3.0~(2\%)}_{-3.5~(3\%)}$ &
$1.18$\\[0.2cm] & $30$ & $ 96.3$ & $^{+24.8~(26\%)}_{-18.4~(19\%)}$ &
$112.8$ & $^{+5.2~(5\%)}_{-7.6~(7\%)}$ & ${}^{+2.8 \, (2\%)}_{-3.3 \,
(3\%)}$ & $1.17$\\[0.2cm] & $35$ & $ 88.1$ &
$^{+22.7~(26\%)}_{-16.8~(19\%)}$ & $102.3$ &
$^{+4.4~(4\%)}_{-6.8~(7\%)}$ & ${}^{+2.5\, (2\%)}_{-2.9 \, (3\%)}$ &
$1.16$\\[0.2cm] & $40$ & $ 79.7$ & $^{+20.7~(26\%)}_{-15.3~(19\%)}$ &
$ 91.8$ & $^{+3.8~(4\%)}_{-6.0~(7\%)}$ & ${}^{+2.3 \, (2\%)}_{-2.6\,
(3\%)}$ & $1.15$\\ &&&&&&&\\ \hline \hline &&&&&&\\ MMHT14 & $25$ &
$102.2$ & $^{+27.0~(26\%)}_{-19.9~(19\%)}$ & $123.1$ &
$^{+5.9~(5\%)}_{-8.4~(7\%)}$ & $^{+2.8~(2\%)}_{-2.5~(2\%)}$ &
$1.20$\\[0.2cm] & $30$ & $ 94.8$ & $^{+25.0~(26\%)}_{-18.5~(19\%)}$
&$113.0$ & $^{+5.0~(4\%)}_{-7.5~(7\%)}$ & ${}^{+2.5 \, (2\%)}_{-2.3 \,
(2\%)}$ & $1.19$\\[0.2cm] & $35$ & $ 86.8$ &
$^{+23.0~(27\%)}_{-16.9~(19\%)}$ & $102.5$ &
$^{+4.3~(4\%)}_{-6.7~(7\%)}$ & ${}^{+2.3 \, (2\%)}_{-2.0\, (2\%)}$ &
$1.18$\\[0.2cm] & $40$ & $ 78.5$ & $^{+20.9~(27\%)}_{-15.3~(20\%)}$ &
$ 91.9$ & $^{+3.7~(4\%)}_{-5.9~(6\%)}$ & ${}^{+2.0\, (2\%)}_{-1.8\,
(2\%)}$ & $1.17$\\ &&&&&&&\\ \hline \hline &&&&&&&\\ NNPDF3.0& $25$ &
$106.9$ & $^{+27.7~(26\%)}_{-20.5~(19\%)}$ & $123.2$ &
$^{+6.3~(5\%)}_{-8.7~(7\%)}$ & $^{+2.1~(2\%)}_{-2.1~(2\%)}$ &
$1.15$\\[0.2cm] & $30$ & $ 99.2$ & $^{+25.8~(26\%)}_{-19.1~(19\%)}$
&$113.1$ & $^{+5.4~(5\%)}_{-7.8~(7\%)}$ & ${}^{+1.9\,(2\%)}_{-1.9\,
(2\%)}$ & $1.14$\\[0.2cm] & $35$ & $ 90.8$ &
$^{+23.7~(26\%)}_{-17.5~(19\%)}$ & $102.6$ &
$^{+4.7~(5\%)}_{-6.8~(7\%)}$ & ${}^{+1.7 \, (2\%)}_{-1.7\, (2\%)}$ &
$1.13$\\[0.2cm] & $40$ & $ 82.1$ & $^{+21.5~(26\%)}_{-15.9~(19\%)}$ &
$ 92.0$ & $^{+4.0~(4\%)}_{-6.1~(7\%)}$ & ${}^{+1.6\, (2\%)}_{-1.6\,
(2\%)}$ & $1.12$\\ &&&&&&&\\ \hline \hline
\end{tabular}
\end{center}
\caption{\label{tab:1} \it LO and NLO integrated fiducial cross
sections for the $pp\to e^+\nu_e\, \mu^-\bar{\nu}_\mu\, e^+\nu_e \,
b\bar{b} +X$ process at the LHC with $\sqrt{s}=13$ TeV. Results are
evaluated using $\mu_R=\mu_F=\mu_0$ with $\mu_0=m_t+m_W/2$. Three PDF
sets and four different values of the $p_T(j_b)$ cut are used. Also
given are theoretical uncertainties coming from the scale variation
$(\delta_{scale})$ and from PDFs $(\delta_{\rm PDF})$.  In the last
column the ${\cal K}$-factor, defined as $\sigma^{\rm NLO}/\sigma^{\rm
LO}$, is shown.}
\end{table}
%

In Table \ref{tab:1} a stability test of LO and NLO fiducial cross
sections with respect to the $b$-jet transverse momentum cut is shown
for $\mu_0=m_t+m_W/2$ and for three PDF sets. The cut is varied in
steps of $5$ GeV within the following range $p_T(j_b) \in (25-40)$
GeV.  We denote theoretical uncertainties as estimated from the scale
variation by $\delta_{\rm scale}$ and from PDFs by $\delta_{\rm
PDF}$. Also given is the ${\cal K}$-factor, defined as ${\cal
K}=\sigma^{\rm NLO}/\sigma^{\rm LO}$. Regardless of the PDF set
employed we observe that NLO QCD corrections are almost constant in
size. Moreover, higher-order theoretical predictions show a very stable
behaviour with respect to theoretical uncertainties. In particular, no
large differences can be observed between the results obtained
for the highest value of the $p_T(j_b)$ cut and for the default value
of $25$ GeV.  This suggests that the perturbative expansion for the
process at hand is not spoiled by the appearance of large logarithms,
thus, under excellent theoretical control. Having established
the stability of the NLO QCD results with respect to the $p_T(j_b)$ cut
for the fixed scale choice we move on to  the
dynamical scale choice, that we have adopted for our studies.

%
\begin{table}[t!]
\begin{center}
  \begin{tabular}{cccccccc}
    \hline \hline
    &&&&&&&\\ 
    \multicolumn{8}{c}{$\mu_R=\mu_F=\mu_0=H_T/3$} \\
    &&&&&&&\\ 
  \hline \hline
  &&&&&&&\\ PDF &$p_{T}(j_b)$ & $\sigma^{\rm LO}$ [ab] &
$\delta_{\rm scale}$ & $\sigma^{\rm NLO}$ [ab] & $\delta_{\rm scale}$ &
$\delta_{\rm PDF}$ & $\sigma^{\rm NLO}/\sigma^{\rm LO}$ \\ &&&&&&&\\
\hline \hline &&&&&&&\\ CT14 & $25$ & $111.7$ &
$^{+29.3~(26\%)}_{-21.6~(19\%)}$ & $124.1$ &
$^{+3.9~(3\%)}_{-7.5~(6\%)}$ & $^{+3.0~(2\%)}_{-3.5~(3\%)}$ &
$1.11$\\[0.2cm] & $30$ & $103.3$ & $^{+27.1~(26\%)}_{-20.0~(19\%)}$ &
$113.5$ & $^{+3.3~(3\%)}_{-6.6~(6\%)}$ & ${}^{+2.8\, (2\%)}_{-3.2\,
(3\%)}$ & $1.10$\\[0.2cm] & $35$ & $ 94.1$ &
$^{+24.7~(26\%)}_{-18.2~(19\%)}$ & $102.7$ &
$^{+3.0~(3\%)}_{-5.9~(6\%)}$ & ${}^{+2.5\, (2\%)}_{-2.9\, (3\%)}$ &
$1.09$\\[0.2cm] & $40$ & $ 84.6$ & $^{+22.3~(26\%)}_{-16.4~(19\%)}$ &
$ 91.9$ & $^{+2.7~(3\%)}_{-5.2~(6\%)}$ & ${}^{+2.2\, (2\%)}_{-2.6\,
(3\%)}$ & $1.09$\\ &&&&&&&\\ \hline \hline &&&&&&&\\ MMHT14 & $25$ &
$110.0$ & $^{+29.6~(27\%)}_{-21.7~(20\%)}$ & $124.3$ &
$^{+3.9~(3\%)}_{-7.4~(6\%)}$ & $^{+2.7~(2\%)}_{-2.4~(2\%)}$ &
$1.13$\\[0.2cm] & $30$ & $101.8$ & $^{+27.4~(27\%)}_{-20.1~(20\%)}$ &
$113.7$ & $^{+3.3~(3\%)}_{-6.6~(6\%)}$ & ${}^{+2.5\, (2\%)}_{-2.2 \,
(2\%)}$ & $1.12$\\[0.2cm] & $35$ & $ 92.8$ &
$^{+25.0~(27\%)}_{-18.3~(20\%)}$ & $102.9$ &
$^{+3.0~(3\%)}_{-5.8~(6\%)}$ & ${}^{+2.3\, (2\%)}_{-2.0\, (2\%)}$ &
$1.11$\\[0.2cm] & $40$ & $ 83.4$ & $^{+22.5~(27\%)}_{-16.5~(20\%)}$ &
$ 92.1$ & $^{+2.7~(3\%)}_{-5.1~(6\%)}$ & ${}^{+2.0\, (2\%)}_{-1.8\,
(2\%)}$ & $1.10$\\ &&&&&&&\\ \hline \hline &&&&&&&\\ NNPDF3.0& $25$ &
$115.1$ & $^{+30.5~(26\%)}_{-22.5~(20\%)}$ & $124.4$ &
$^{+4.3~(3\%)}_{-7.7~(6\%)}$ & $^{+2.1~(2\%)}_{-2.1~(2\%)}$ &
$1.08$\\[0.2cm] & $30$ & $106.5$ & $^{+28.2~(26\%)}_{-20.8~(20\%)}$ &
$113.9$ & $^{+3.5~(3\%)}_{-6.8~(6\%)}$ & ${}^{+1.9\, (2\%)}_{-1.9\,
(2\%)}$ & $1.07$\\[0.2cm] & $35$ & $ 97.0$ &
$^{+25.7~(27\%)}_{-18.9~(20\%)}$ & $103.1$ &
$^{+3.1~(3\%)}_{-6.0~(6\%)}$ & ${}^{+1.7\, (2\%)}_{-1.7\, (2\%)}$ &
$1.06$\\[0.2cm] & $40$ & $ 87.2$ & $^{+23.2~(27\%)}_{-17.0~(20\%)}$ &
$ 92.3$ & $^{+2.8~(3\%)}_{-5.3~(6\%)}$ & ${}^{+1.5\, (2\%)}_{-1.5\,
(2\%)}$ & $1.06$\\ &&&&&&&\\ \hline
  \hline
\end{tabular}
\end{center}
\caption{\label{tab:2} \it  As in Table \ref{tab:1} but 
    for $\mu_R=\mu_F=\mu_0=H_T/3$.}
\end{table}
%
%
\begin{figure}[t!]
\begin{center}
  \includegraphics[width=0.49\textwidth]{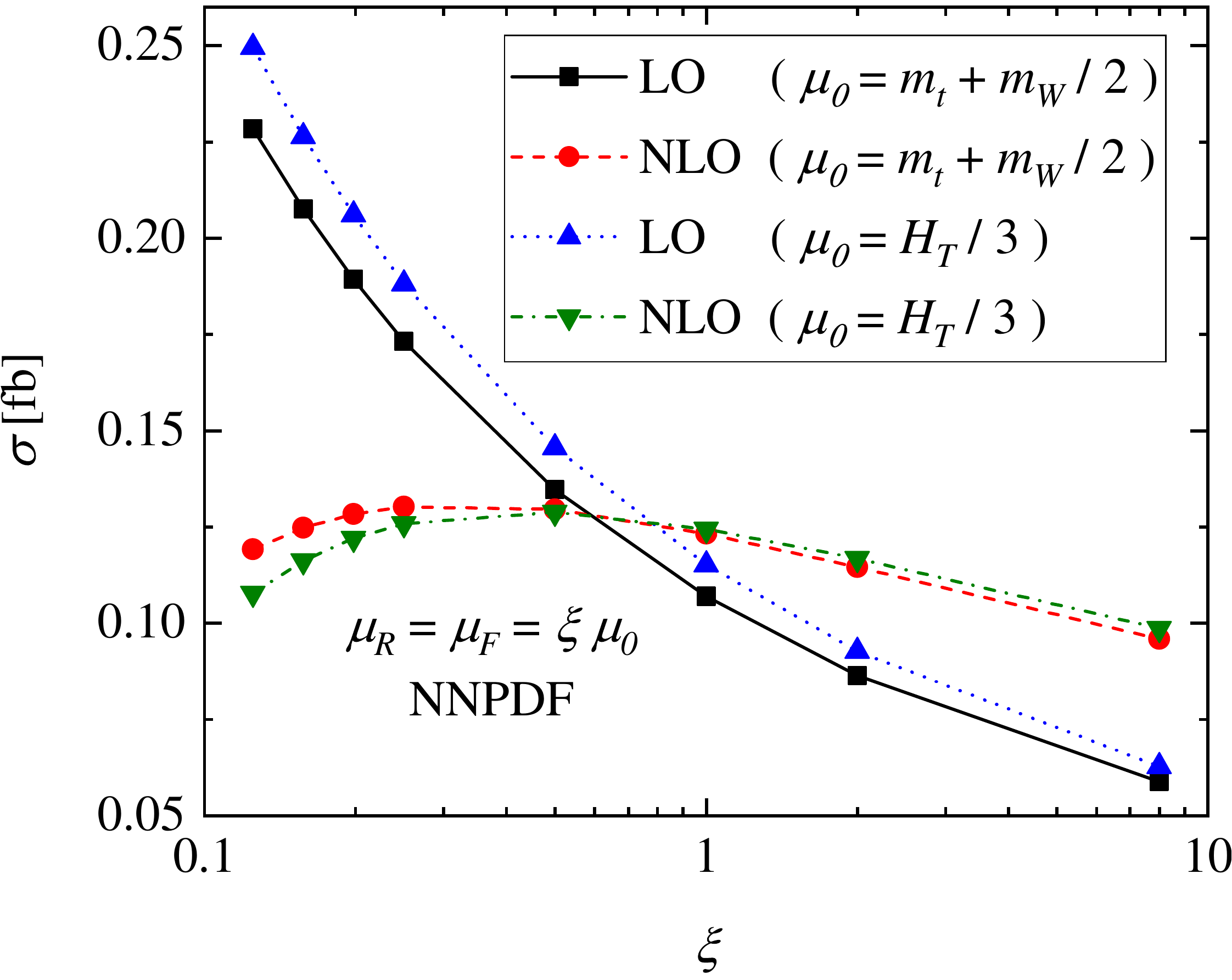}
\end{center}
\caption{\it Scale dependence of the LO and NLO integrated fiducial cross
section for the $pp\to e^+\nu_e\, \mu^-\bar{\nu}_\mu\, e^+\nu_e \,
b\bar{b} +X$ production process at the LHC with $\sqrt{s}=13$
TeV. Renormalisation and factorisation scales are set to the common
value $\mu_R=\mu_F=\mu_0$ with $\mu_0=m_t+m_W/2$ and
$\mu_0=H_T/3$. The LO and NLO NNPDF3.0 PDF sets are employed.}
\label{fig:scale1}
\end{figure}
%
%
\begin{figure}[h!]
  \begin{center}
    \vspace{0.5cm}
  \includegraphics[width=0.49\textwidth]{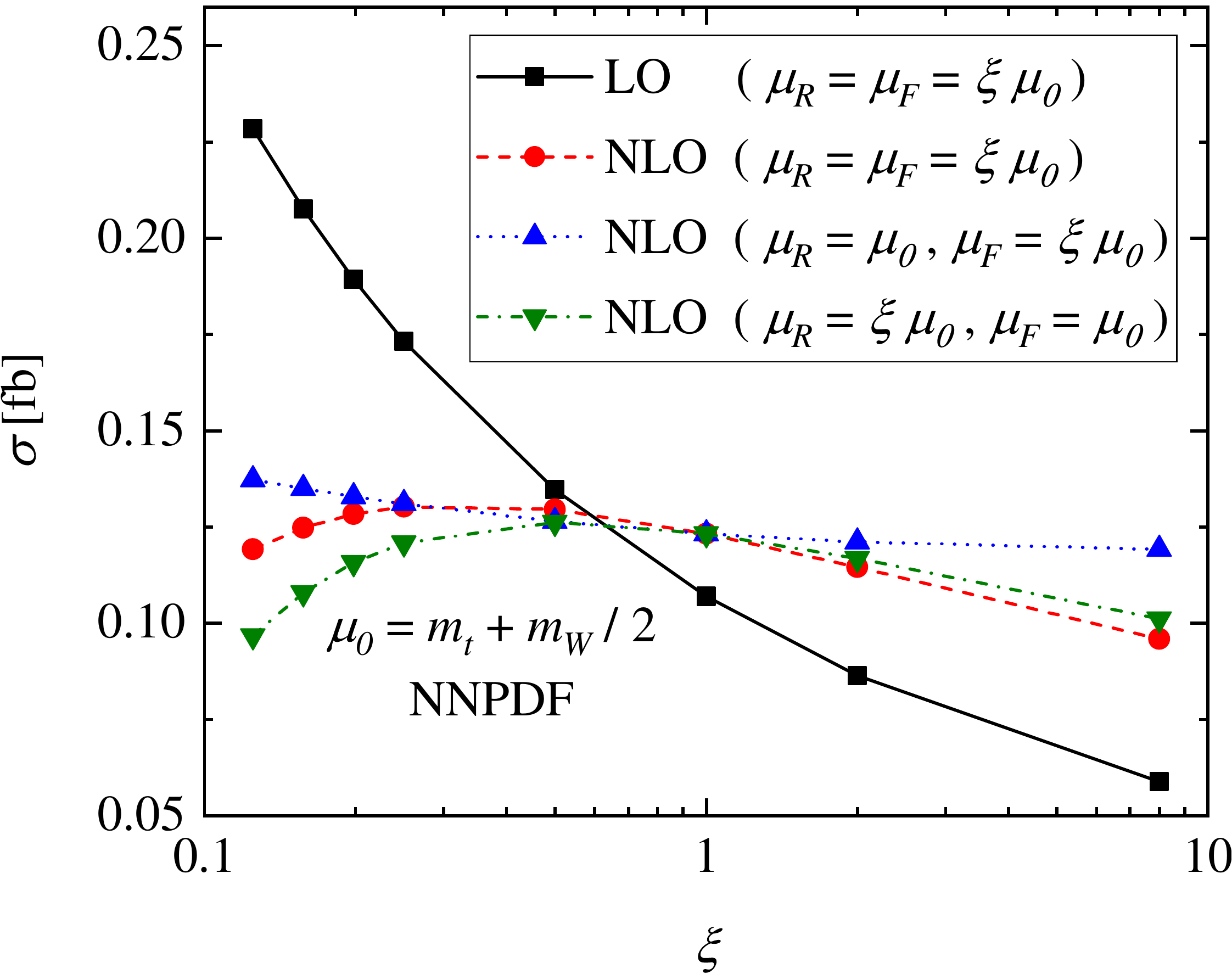}
  \includegraphics[width=0.49\textwidth]{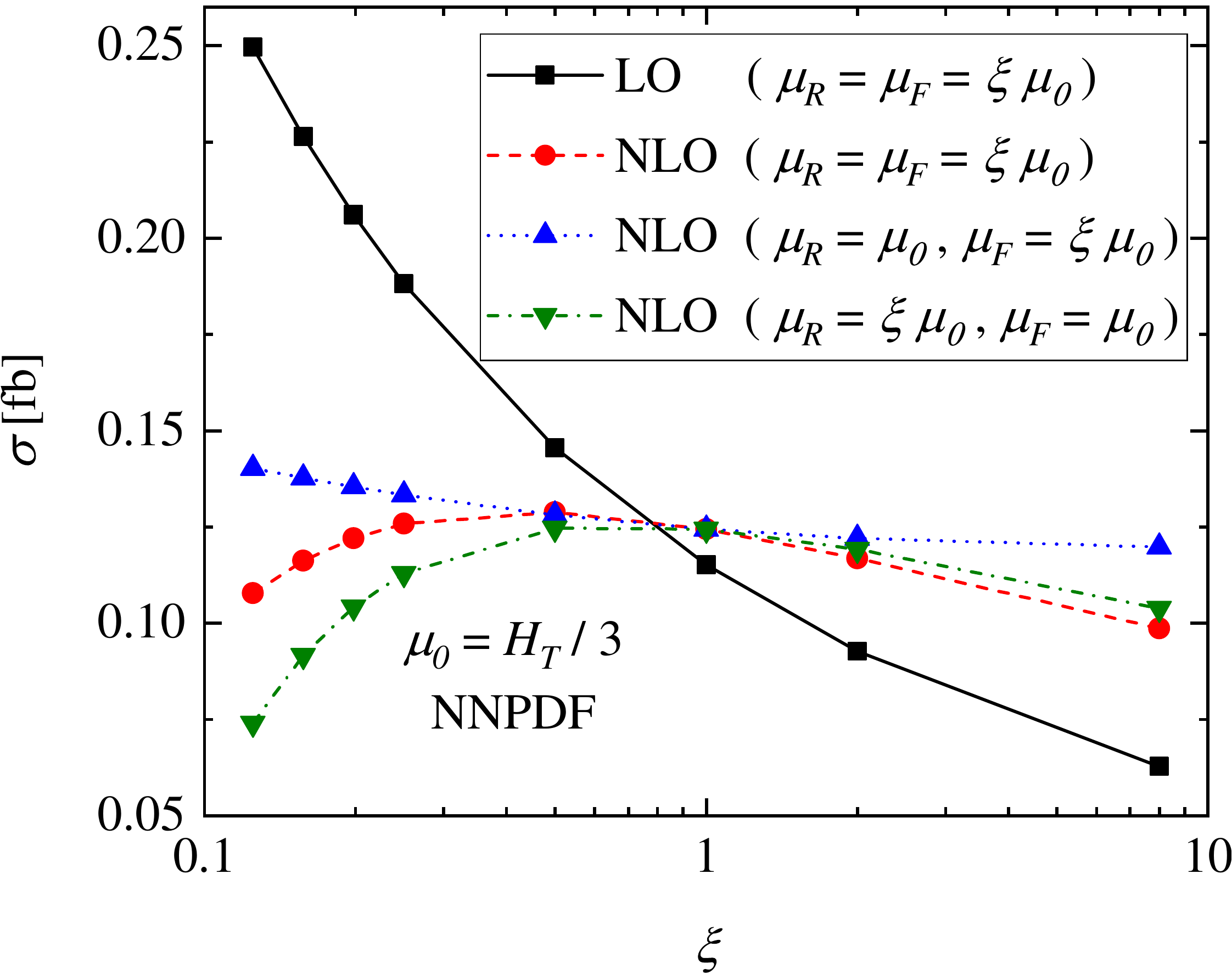}
\end{center}
\caption{\it Scale dependence of the LO and NLO integrated fiducial cross
section for the $pp\to e^+\nu_e\, \mu^-\bar{\nu}_\mu\, e^+\nu_e \,
b\bar{b} +X$ production process at the LHC with $\sqrt{s}=13$
TeV separately for  $\mu_0=m_t+m_W/2$ and
$\mu_0=H_T/3$. The LO and NLO NNPDF3.0 PDF sets are employed. For each
case of $\mu_0$ also shown is the variation of $\mu_R$ with fixed
$\mu_F$ and the variation of $\mu_F$ with fixed $\mu_R$.}
\label{fig:scale2}
\end{figure}
%
%
\begin{figure}[t]
\begin{center}
\includegraphics[width=0.49\textwidth]{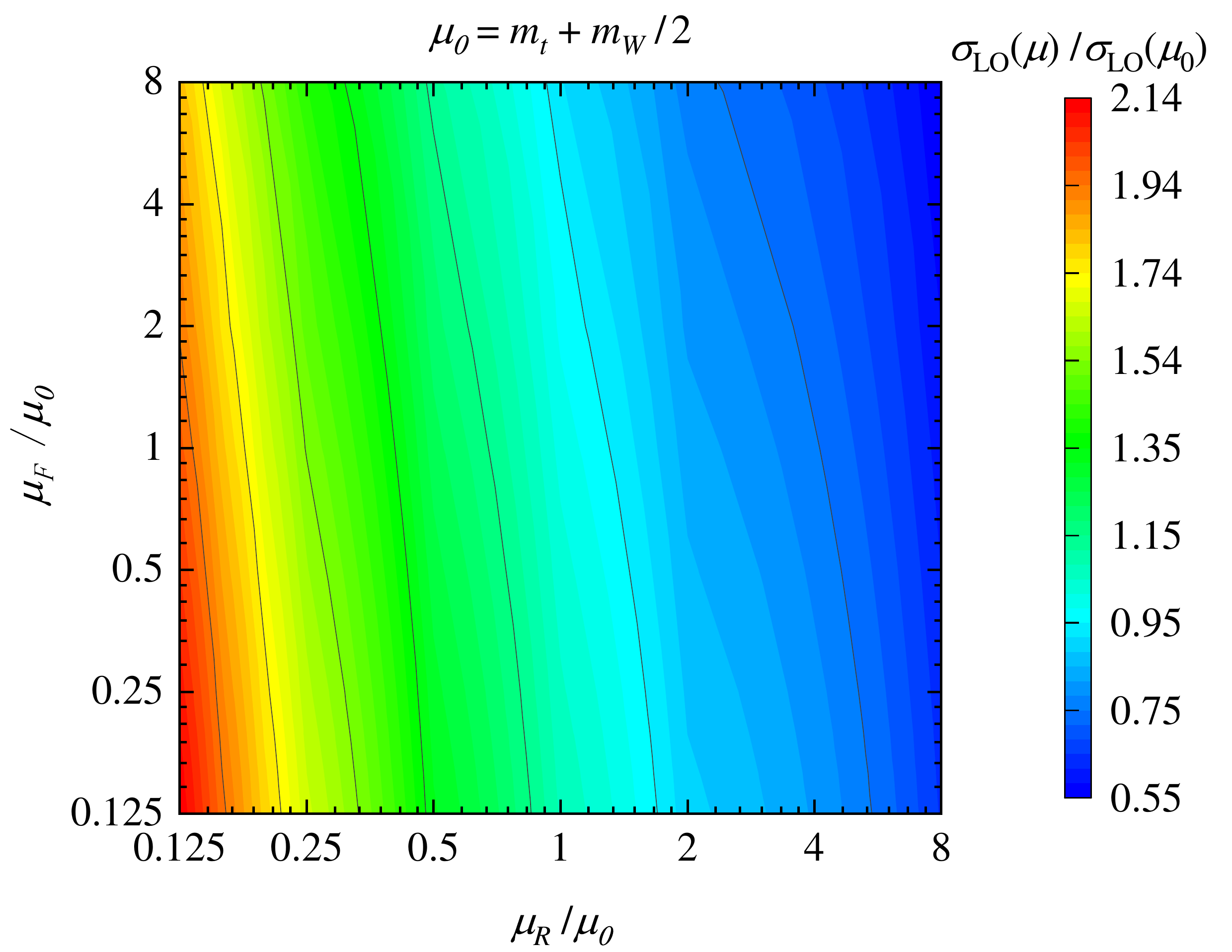}
\includegraphics[width=0.49\textwidth]{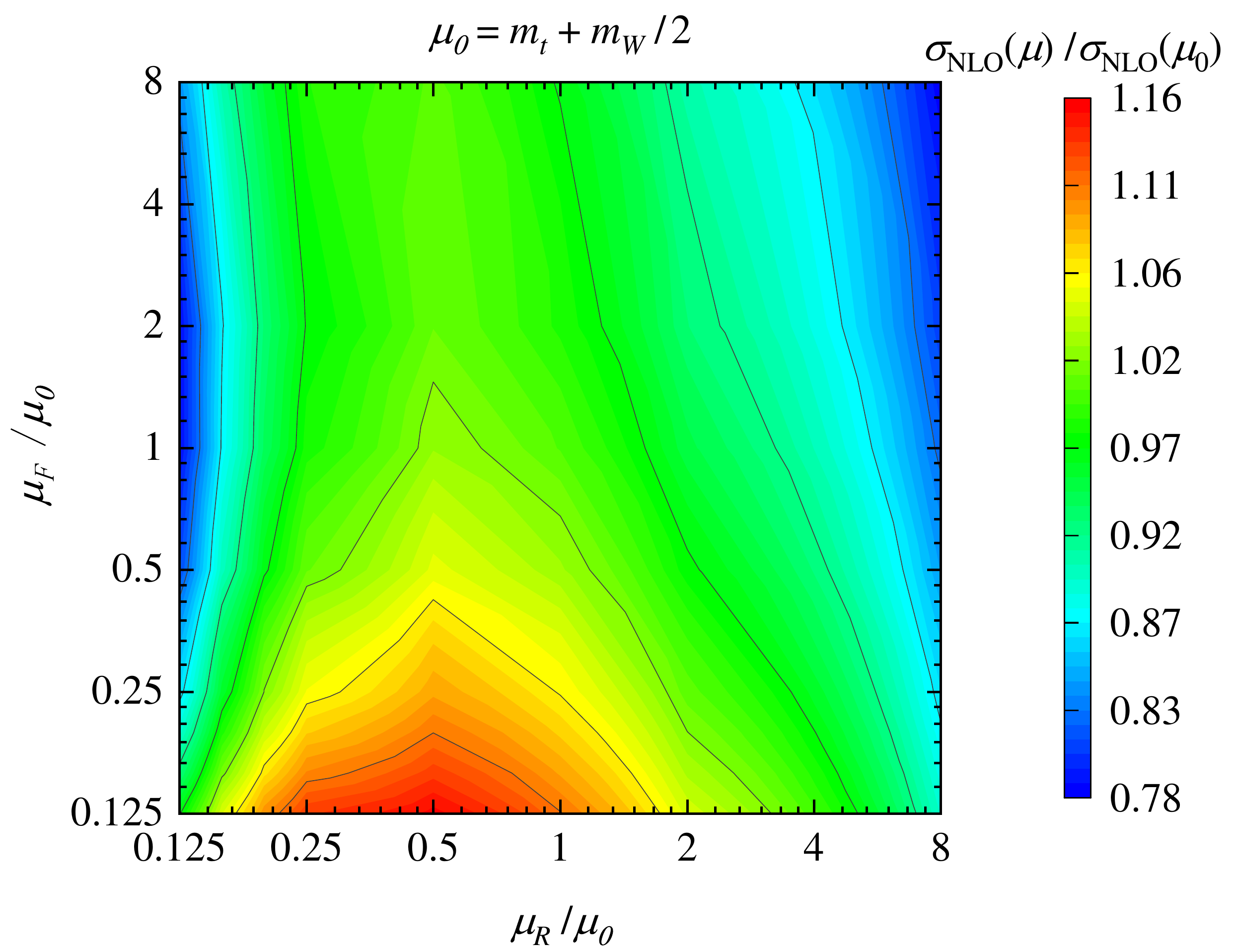}\\
\vspace{0.4cm}
\includegraphics[width=0.49\textwidth]{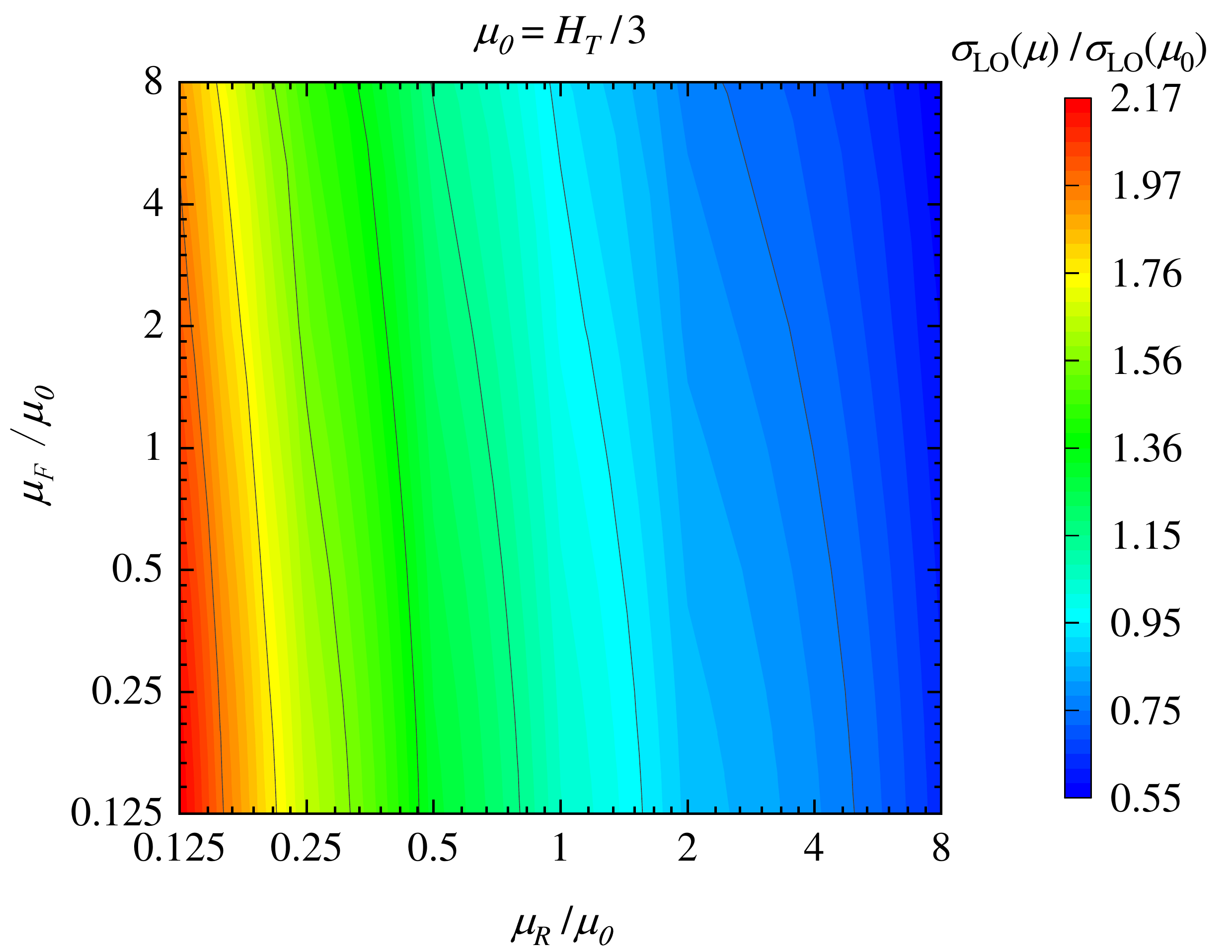}
\includegraphics[width=0.49\textwidth]{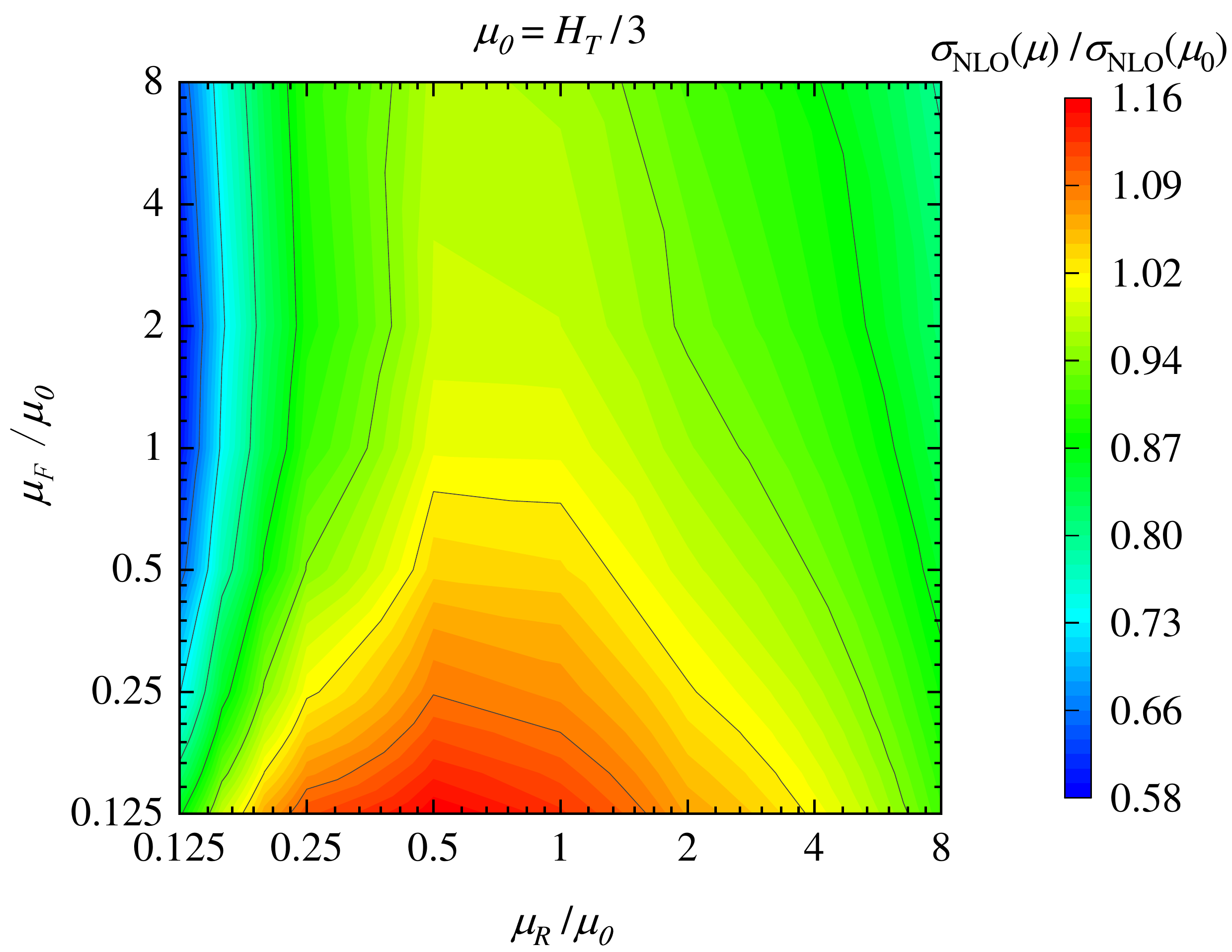}
\end{center}
\caption{\it Integrated fiducial cross section for the $pp\to e^+\nu_e\,
\mu^-\bar{\nu}_\mu\, e^+\nu_e \, b\bar{b} +X$ process at the LHC with
$\sqrt{s}=13$ TeV as a function of $\mu_R$ and $\mu_F$. Results are
evaluated using $\mu_0=m_t+m_W/2$ and $\mu_0=H_T/3$. The LO and NLO
NNPDF3.0 PDF sets are employed.}
\label{fig:scale3}
\end{figure}
%
%

Using the same input parameters and cuts  as before but employing
$\mu_0=H_T/3$ the results for  the $pp \to e^+\nu_e\,
\mu^-\bar{\nu}_\mu\, e^+\nu_e \, b\bar{b}$ process for the default
NNPDF3.0 PDF sets can be summarised as follows
\begin{equation}
  \begin{split}
&\sigma^{\rm LO}_{e^+\nu_e\,
\mu^-\bar{\nu}_\mu\, e^+\nu_e \, b\bar{b}} \left({\rm NNPDF3.0},
\mu_0=H_T/3\right)= 115.1^{\,+30.5\, (26\%)}_{\,-22.5\, (20\%)}
\, {\rm [scale]} \, {\rm ab}\,,
\\[0.2cm]
&\sigma^{\rm NLO}_{e^+\nu_e\, 
\mu^-\bar{\nu}_\mu\, e^+\nu_e \, b\bar{b}} \left({\rm NNPDF3.0},
\mu_0= H_T/3\right)= 124.4^{\,+4.3~(3\%)}_{\,-7.7~(6\%)}
\, {\rm
[scale]} \, {}^{+2.1~(2\%)}_{-2.1~(2\%)}\, {\rm [PDF]} \, {\rm ab}\,.
\end{split}
\end{equation}  
For the MMHT14 PDF sets, on the other hand, we have
\begin{equation}
  \begin{split}
&\sigma^{\rm LO}_{e^+\nu_e\,
\mu^-\bar{\nu}_\mu\, e^+\nu_e \, b\bar{b}} \left({\rm MMHT14},
\mu_0=H_T/3\right)=   110.0^{\,+29.6\, (27\%)}_{\,-21.7\, (20\%)}
\, {\rm [scale]} \, {\rm ab}\,,
\\[0.2cm]
&\sigma^{\rm NLO}_{e^+\nu_e\, 
\mu^-\bar{\nu}_\mu\, e^+\nu_e \, b\bar{b}} \left({\rm MMHT14},
\mu_0= H_T/3\right)=  124.3^{\,+3.9\, (3\%)}_{\,-7.4\,(6\%)}
\, {\rm
[scale]} \, {}^{+2.7\, (2\%)}_{-2.4\, (2\%)}   \, {\rm [PDF]} \, {\rm ab}\,.
\end{split}
\end{equation}  
Lastly, for the CT14 PDF sets we can report on the following predictions  
\begin{equation}
  \begin{split}
&\sigma^{\rm LO}_{e^+\nu_e\,
\mu^-\bar{\nu}_\mu\, e^+\nu_e \, b\bar{b}} \left({\rm CT14},
\mu_0=H_T/3\right)=   111.7^{\,+29.3\, (26\%)}_{\,-21.6\,(19\%)}
\, {\rm [scale]} \, {\rm ab}\,,
\\[0.2cm]
&\sigma^{\rm NLO}_{e^+\nu_e\, 
\mu^-\bar{\nu}_\mu\, e^+\nu_e \, b\bar{b}} \left({\rm CT14},
\mu_0= H_T/3\right)=  124.1^{\,+3.9\, (3\%)}_{\,-7.5 \, (6\%)}
\, {\rm
[scale]} \, {}^{+3.0\, (2\%)}_{-3.5 \, (3\%)}  \, {\rm [PDF]} \, {\rm ab}\,.
\end{split}
\end{equation}
Results with $\mu_0=H_T/3$ and for the NNPDF3.0 PDF set
are a bit higher, as they increased by $8\%$ at LO and by $1\%$ at NLO
when compared with results from Table \ref{tab:1}. This is perfectly
within the theoretical error estimates at the corresponding
perturbative order. Moreover, the ${\cal K}$-factor obtained with this
new scale is smaller, of the order of ${\cal K}=1.08$. This is the
consequence of the larger shift in the normalisation of the LO cross
section, which depends more strongly on the changes in $\mu_R$ and
$\mu_F$. The size of the NLO QCD corrections is rather stable and
increases up to $11\%-13\%$ for CT14 and MMHT14 respectively. As for
the theoretical uncertainties from the scale dependence and from PDFs
they are at the same level as for the fixed scale choice. Differences
between predictions for various PDF sets are of the order of $3\%-5\%$
at LO and $0.1\%-0.2\%$ at NLO. Thus, internal PDF uncertainties as
calculated separately for NNPDF3.0, MMHT14 and CT14 are an order of
magnitude larger. Still, uncertainties due to scale dependence are the
dominant source of theoretical systematics.

The integrated LO and NLO fiducial cross sections for the $pp\to
e^+\nu_e\, \mu^-\bar{\nu}_\mu\, e^+\nu_e \, b\bar{b} +X$ production
process for the dynamical scale choice are shown in Table \ref{tab:2}
for four different values of the $p_T(j_b)$ cut. Also for
$\mu_0=H_T/3$ we observe a very stable behaviour of the cross section
with respect to the higher-order corrections.  Moreover, theoretical
uncertainties do not show any sensitivity to changes in the $p_T(j_b)$
cut value.

In Figure \ref{fig:scale1} we present the result for the scale
dependence graphically for $\mu_0=m_t+m_W/2$ and $\mu_0=H_T/3$. The
behaviour of LO and NLO cross sections for the default NNPDF3.0 PDF
sets is presented upon varying the $\mu_R$ and $\mu_F$ scales
simultaneously by a factor $\xi$ in the following range $\xi \in
\left\{0.125, . . . , 8\right\}$. As already discussed, at LO the
dependence is large illustrating the well known fact that the LO
prediction can only provide a rough estimate. A significant reduction
in the scale uncertainty is observed when NLO QCD corrections are
included.

In Figure \ref{fig:scale2} we display again the dependence of the
integrated LO and NLO fiducial cross sections on the variation of the fixed and
dynamical scales for the NNPDF3.0 PDF set. This time, however, we show
additionally NLO results with individual variation of $\mu_R$ and
$\mu_F$. Each time we plot two extra curves, the first one corresponds
to the case where $\mu_R$ is kept fixed at the central value, while
$\mu_F$ is varied and the second one describes the opposite
situation. We can observe that regardless of the scale choice the
scale variation is due to changes in both $\mu_R$ and $\mu_F$. Thus,
it is not driven solely by the renormalisation scale.

The dependence of the LO and NLO cross sections on $\mu_R$ and
$\mu_F$, which are varied this time independently but simultaneously
around a central value of the scale, is presented in Figure
\ref{fig:scale3}. We plot distributions of the LO and NLO cross
sections in the $\mu_R-\mu_F$ plane. On top of the previous three
special cases i) $\mu_R=\mu_F=\xi \mu_0$, ii) $\mu_R=\mu_0$,
$\mu_F=\xi\mu_0$ and iii) $\mu_F=\mu_0$, $\mu_R=\xi\mu_0$, here all
cases in between are depicted as well.  These contour plots provide
complementary information to the previous scale dependence plots. We
can see that at LO, independently of the scale choice, the fiducial
cross section decreases only mildly with increasing $\mu_F$, while it
decreases rapidly with the increment of $\mu_R$. Thus, the LO cross
section dependence on $\mu_R$ is much larger than on $\mu_F$. At NLO
the situation is slightly   different, because the dependence of the
cross section on $\mu_F$ increases substantially. However,
$\sigma^{\rm NLO}$ is still dominated by the changes in $\mu_R$.

%
\subsection{Differential distributions}
%

An important task of studies on higher-order corrections is to examine
how much they can affect the shape of various kinematic
distributions.  It is equally important to estimate the final
theoretical error for differential cross sections. In the following we
shall examine various observables that are of interest for the LHC.
For the default PDF set we plot each observable twice, once for
$\mu_0=m_t+m_W/2$ and once for $\mu_0=H_T/3$. The upper panel of each
plot shows the absolute prediction at LO and NLO together with their
scale dependence bands calculated according to Eq.~\eqref{scan}.  The
lower panels display the same LO and NLO predictions normalised to the
LO result at $\mu_R=\mu_F=\mu_0$.  The blue band provides the relative
scale uncertainty of the LO cross section, whereas the red band gives
the differential ${\cal K}-$factor together with its uncertainty
band. We have examined about $30$ observables. In the following we
shall present, however, just a few examples to highlight the main
features and importance of higher-order QCD corrections for the $pp
\to e^+ \nu_e \, \mu^- \bar{\nu}_\mu \, e^+ \nu_e \, b\bar{b} +X$
process.

\begin{figure}[t!]
\begin{center}
  \includegraphics[width=0.49\textwidth]{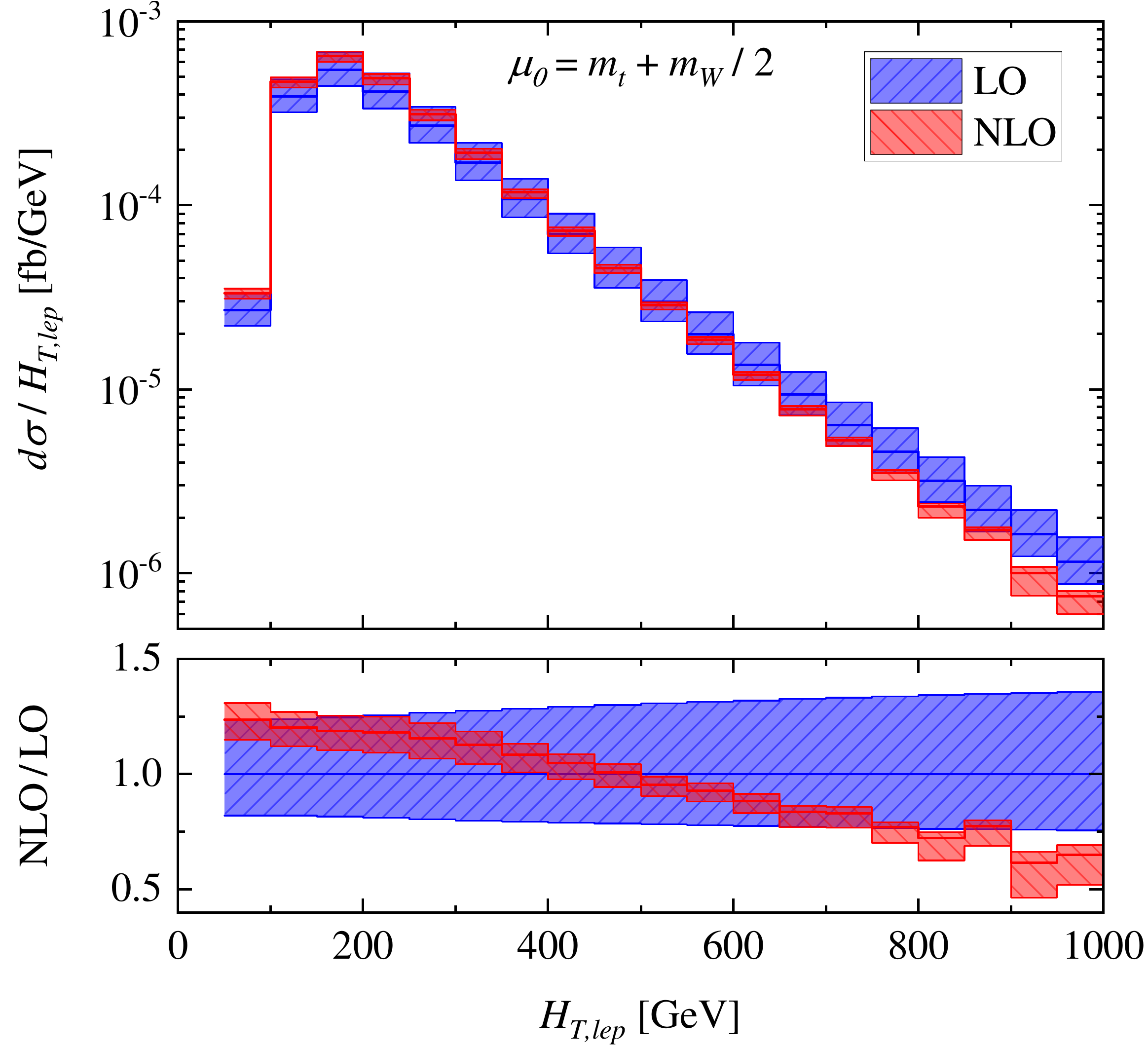}
  \includegraphics[width=0.49\textwidth]{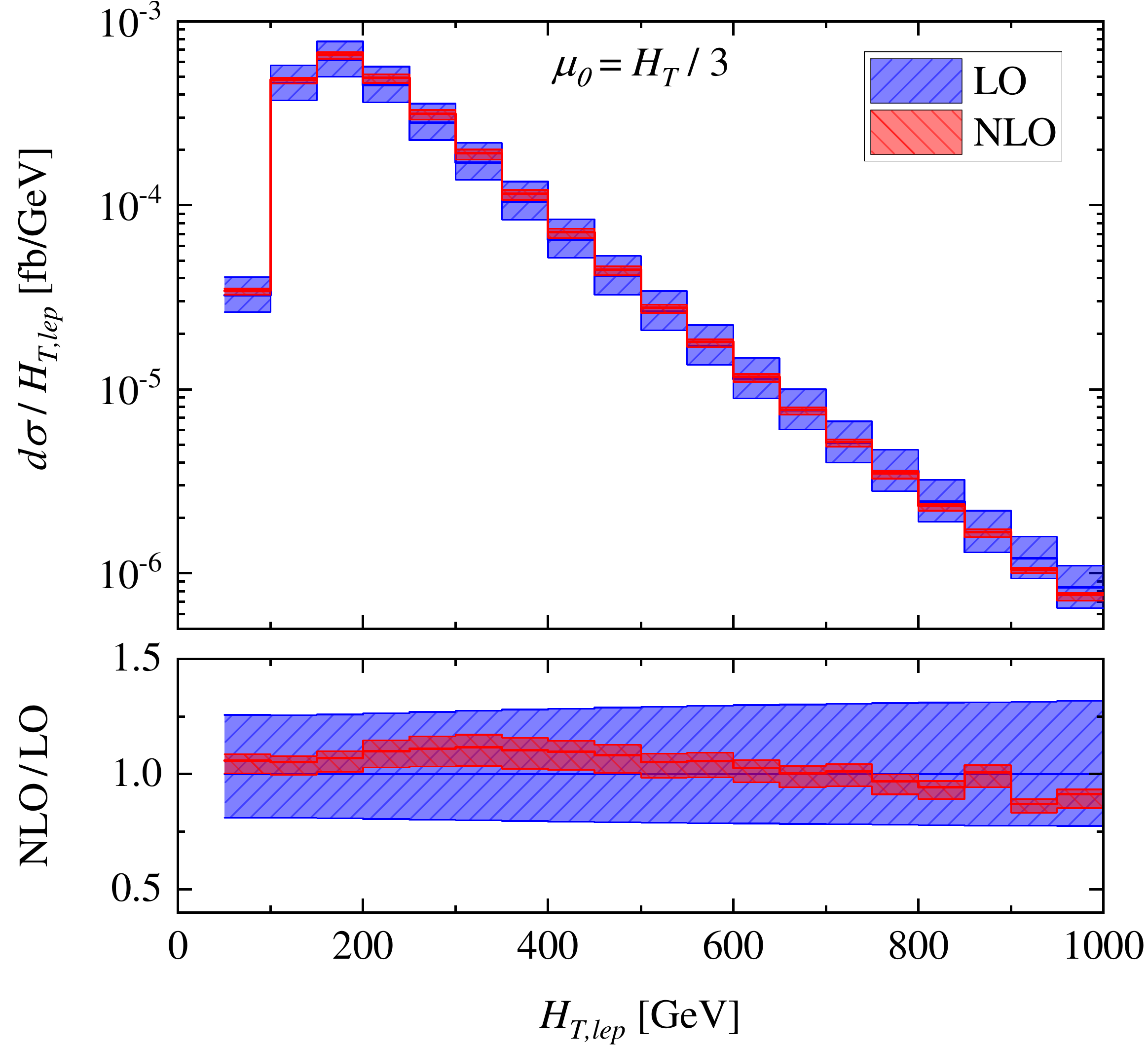}
  \\
\vspace{0.4cm}
  \includegraphics[width=0.49\textwidth]{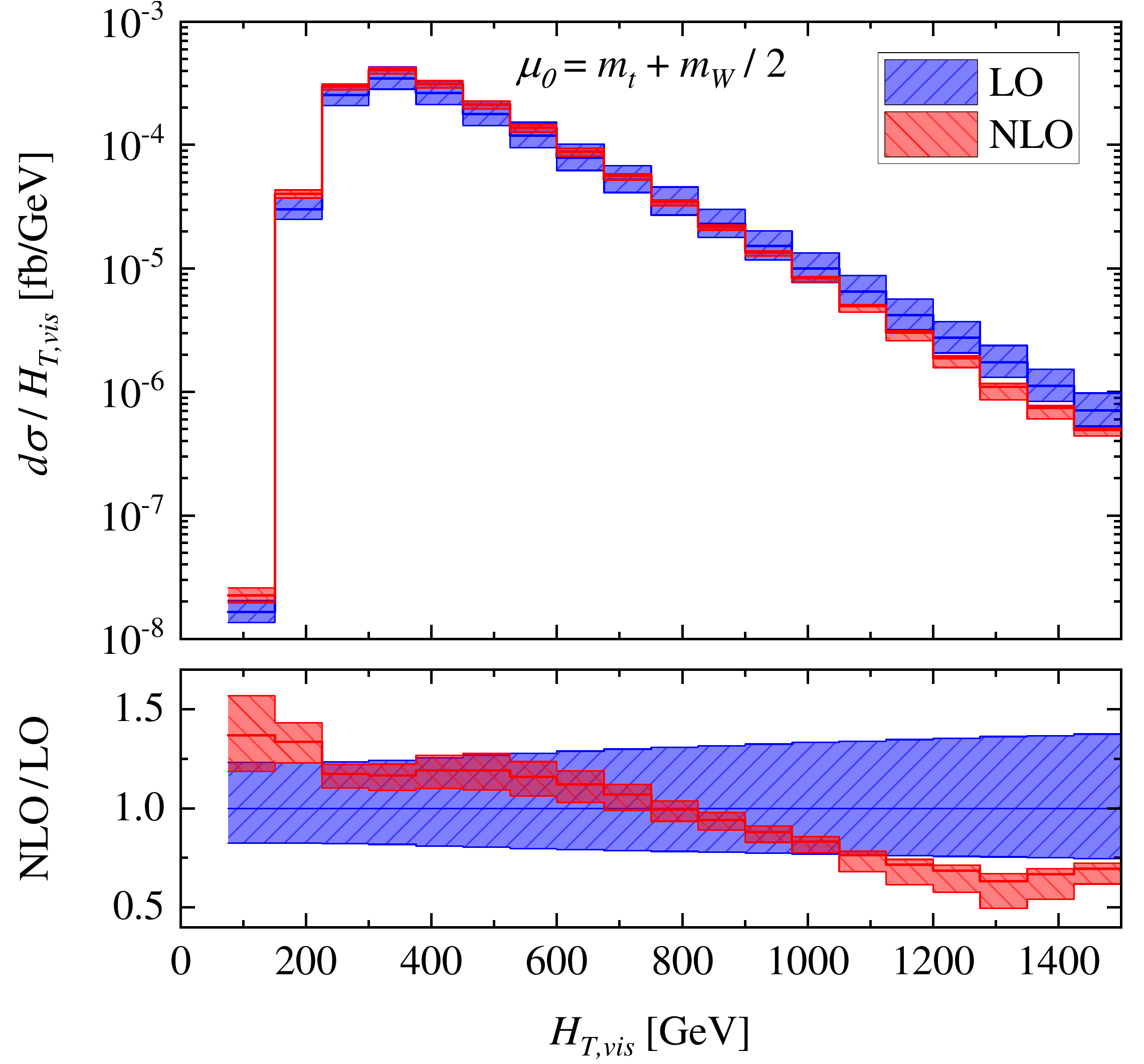}
  \includegraphics[width=0.49\textwidth]{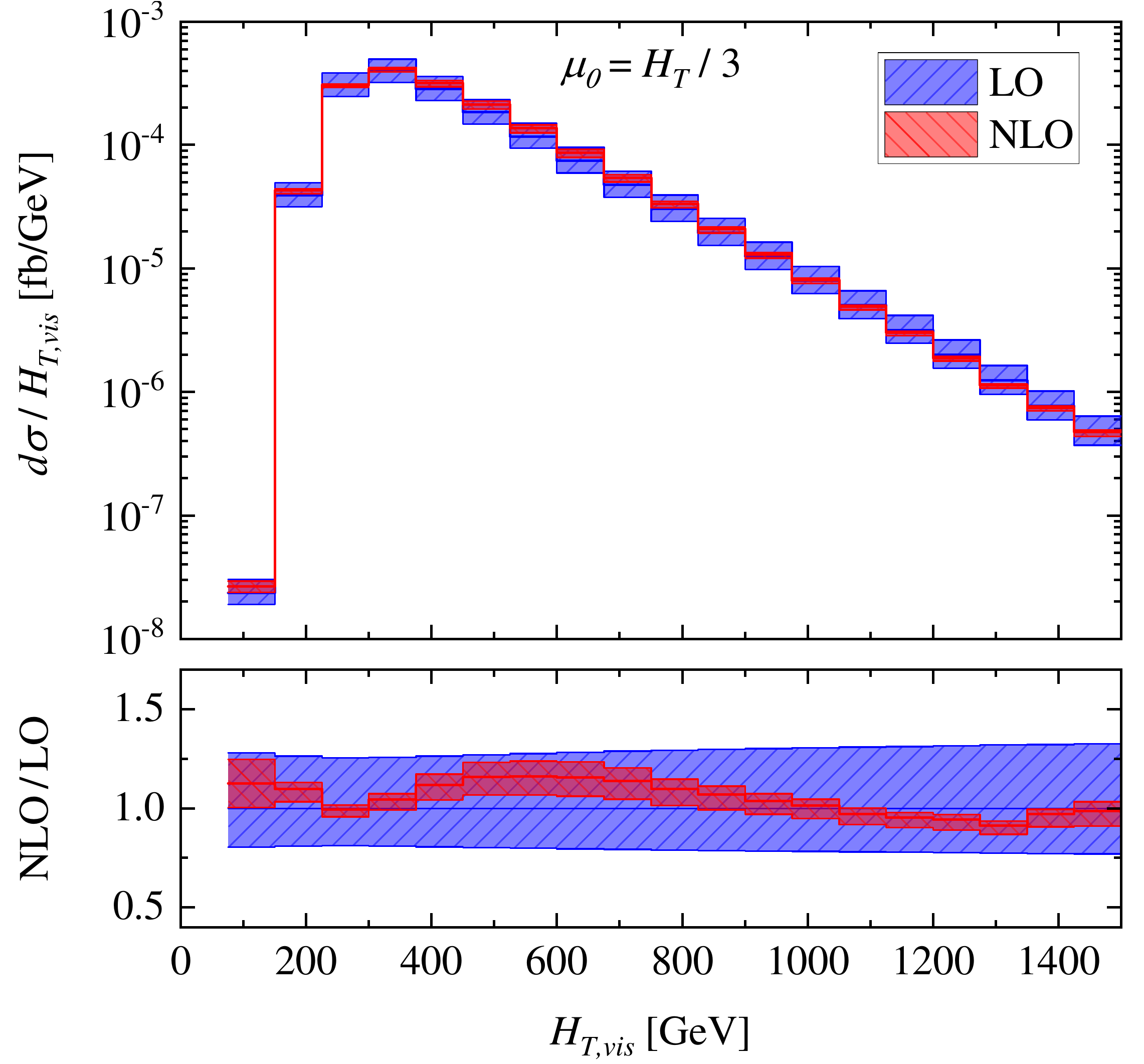}
 \end{center}
 \caption{\it  The $pp\to e^+\nu_e\, \mu^-\bar{\nu}_\mu\,
e^+\nu_e \, b\bar{b} +X$ differential cross section distribution at
the LHC with $\sqrt{s}=13$ TeV as a function of $H_T^{lep}$ and
$H_T^{vis}$.  The upper panels show absolute LO and NLO predictions
together with corresponding uncertainty bands. The lower panels
display the differential ${\cal K}$-factor together with the
uncertainty band and the relative scale uncertainties of the LO cross
section. Results are evaluated using $\mu_R=\mu_F=\mu_0$ with
$\mu_0=m_t+m_W/2$ and $\mu_0=H_T/3$.  The LO and the NLO
NNPDF3.0 PDF  sets are employed. }
\label{fig:diff1}
\end{figure}

 We start with the scalar sum of the transverse momenta
of the charged leptons available in this process, which we label
$H_T^{lep}$ and define as
\begin{equation} H_T^{lep} = p_T(\mu^-) +p_T(\ell_1) + p_T(\ell_2) \,,
\end{equation}
where $\ell_{1,2} = e^+_{1,2}$.  Also examined is the
scalar sum of the transverse momenta of the visible final states
denoted as $H_T^{vis}$ and given by
 \begin{equation} H_T^{vis} = p_T(\mu^-) +p_T(\ell_1) + p_T(\ell_2) +
p_T(j_{b_1}) + p_T(j_{b_2})\,.
\end{equation}
Both observables, which are often exploited in various SM measurements
and BSM searches by the ATLAS and CMS collaborations, are displayed in
Figure \ref{fig:diff1}. We examine first the size of NLO QCD
corrections. With the fixed scale choice higher-order corrections from
about $+25\%$ at the begin of the spectrum down to about $-35\%$ for
the high $p_T$ tails can be observed for $H_T^{lep}$ causing
distortions up to $60\%$.  For the dynamical scale choice, on the
other hand, NLO QCD corrections are up to about $\pm 10\%$ only,
leading to maximal distortions of the order of $20\%$. Still,
independently of the scale choice the $H_T^{lep}$ differential ${\cal
K}$-factor is not flat highlighting the importance of NLO QCD
corrections. A very similar conclusion could be drawn for the second
observable $H_T^{vis}$.  Furthermore, alike for the
$H_T^{lep}$ and $H_T^{vis}$ observables with the fixed scale choice
the NLO error bands do not fit within the LO ones. A scale variation
procedure is considered good only if the error estimate at the LO
contains the central value of the next higher order, see
e.g. \cite{Czakon:2016dgf}, which is not the case here. Thus,
$\mu_0=m_t+m_W/2$ leads to perturbative instabilities in the TeV
region of the differential cross section distributions. The dynamical
scale choice, however, stabilises the tails and keeps the NLO
uncertainties bands within the LO ones as one would expect from a well
behaved perturbative expansion.  As for the theoretical uncertainties
due to the scale dependence also at the differential level we notice a
substantial reduction of the uncertainties when the higher-order QCD
corrections are incorporated.  For both observables theoretical
uncertainties for the fixed scale choice are maximally up to
$15\%-20\%$, whereas for the dynamical scale choice they are of the
order of $5\%-10\%$.
%
%
%
%
\begin{figure}[t!]
  \begin{center}
  \includegraphics[width=0.49\textwidth]{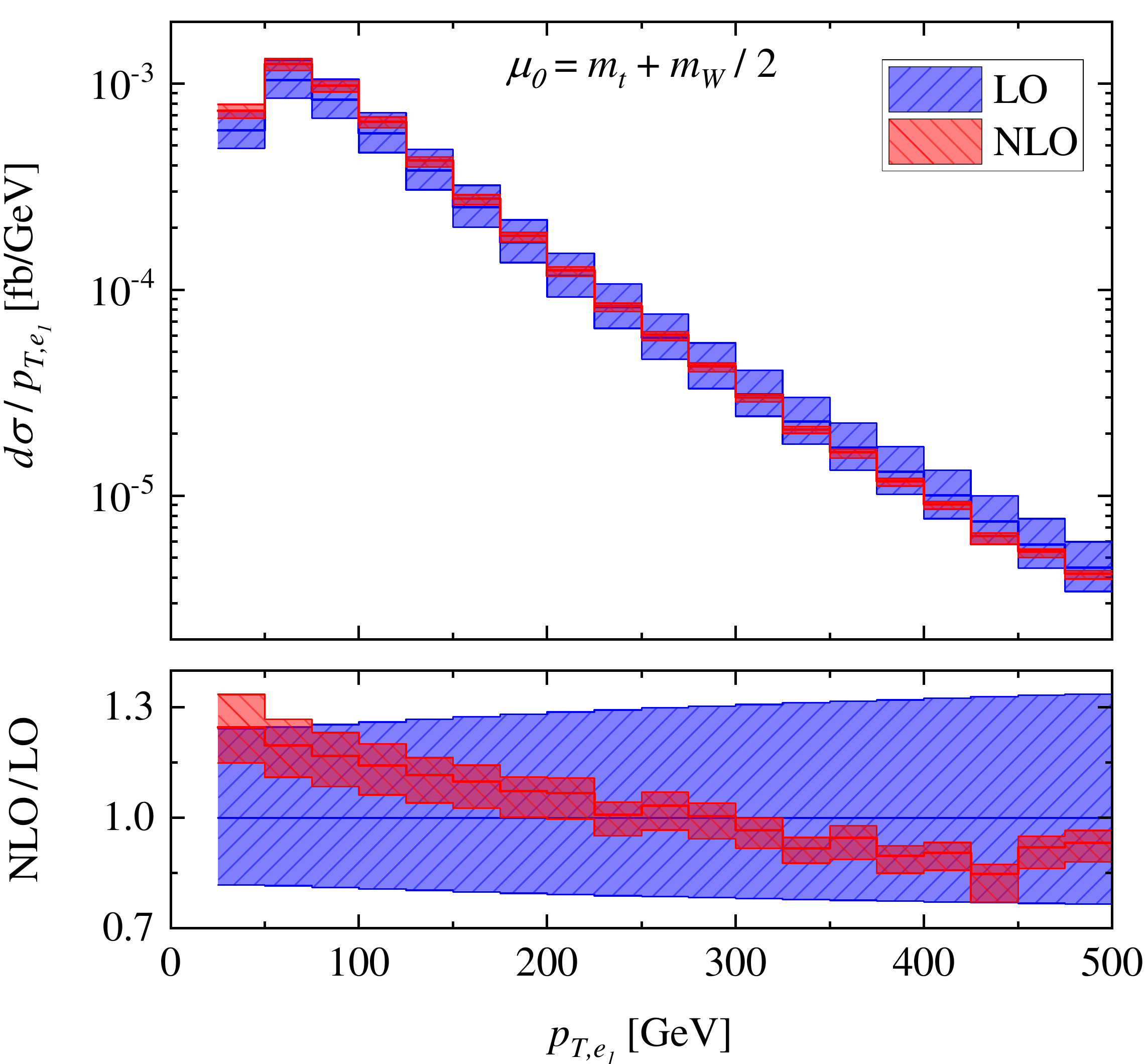}
  \includegraphics[width=0.49\textwidth]{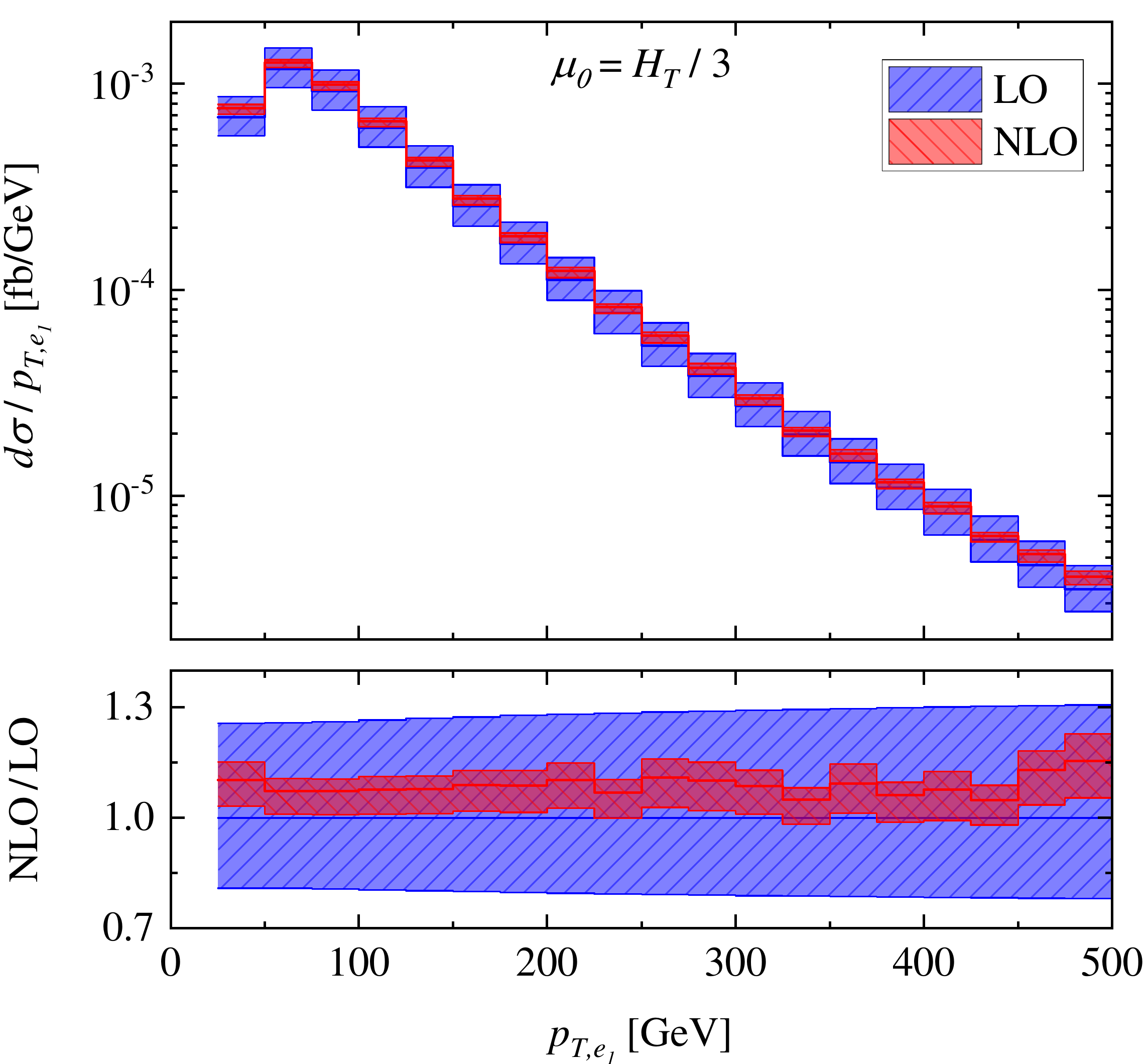}
  \\
\vspace{0.4cm}
     \includegraphics[width=0.49\textwidth]{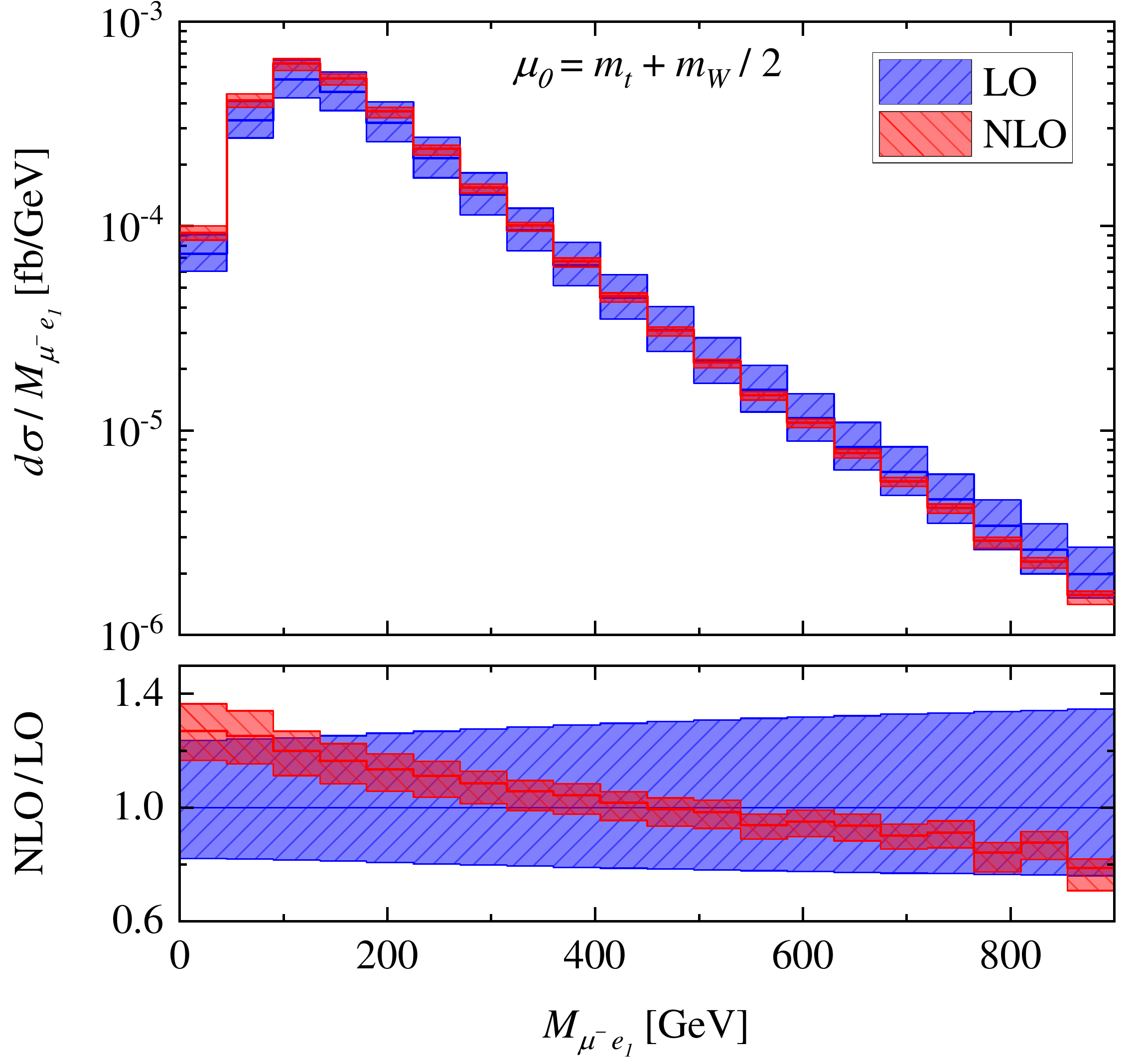}
   \includegraphics[width=0.49\textwidth]{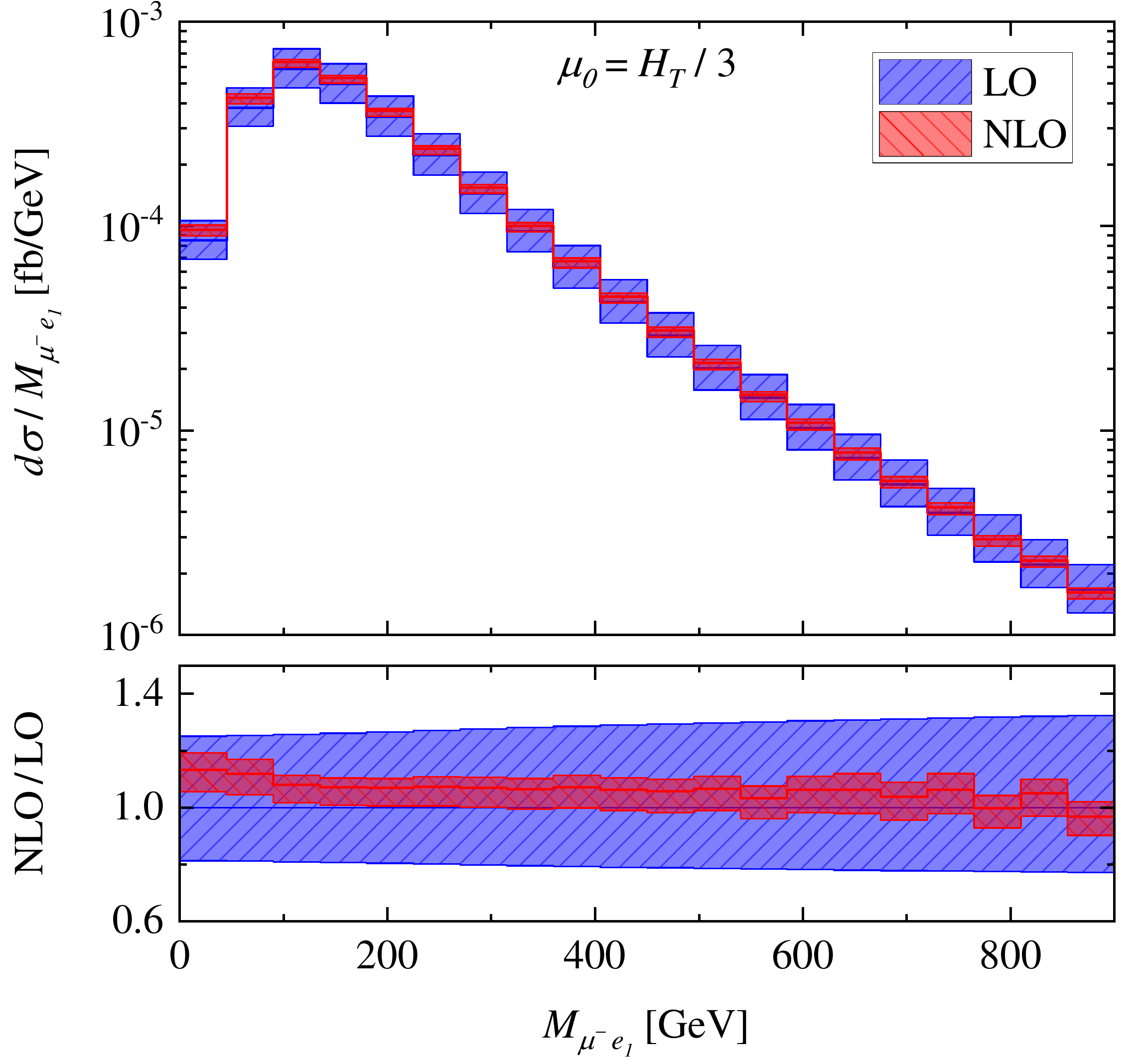}
\end{center}
\caption{\it  As in Figure \ref{fig:diff1} but for the
    $p_{T,\, e_1}$ and $M_{\mu^- e_1}$ distributions.} 
\label{fig:diff2}
\end{figure}
%
%
\begin{figure}[t!]
\begin{center}
   \includegraphics[width=0.49\textwidth]{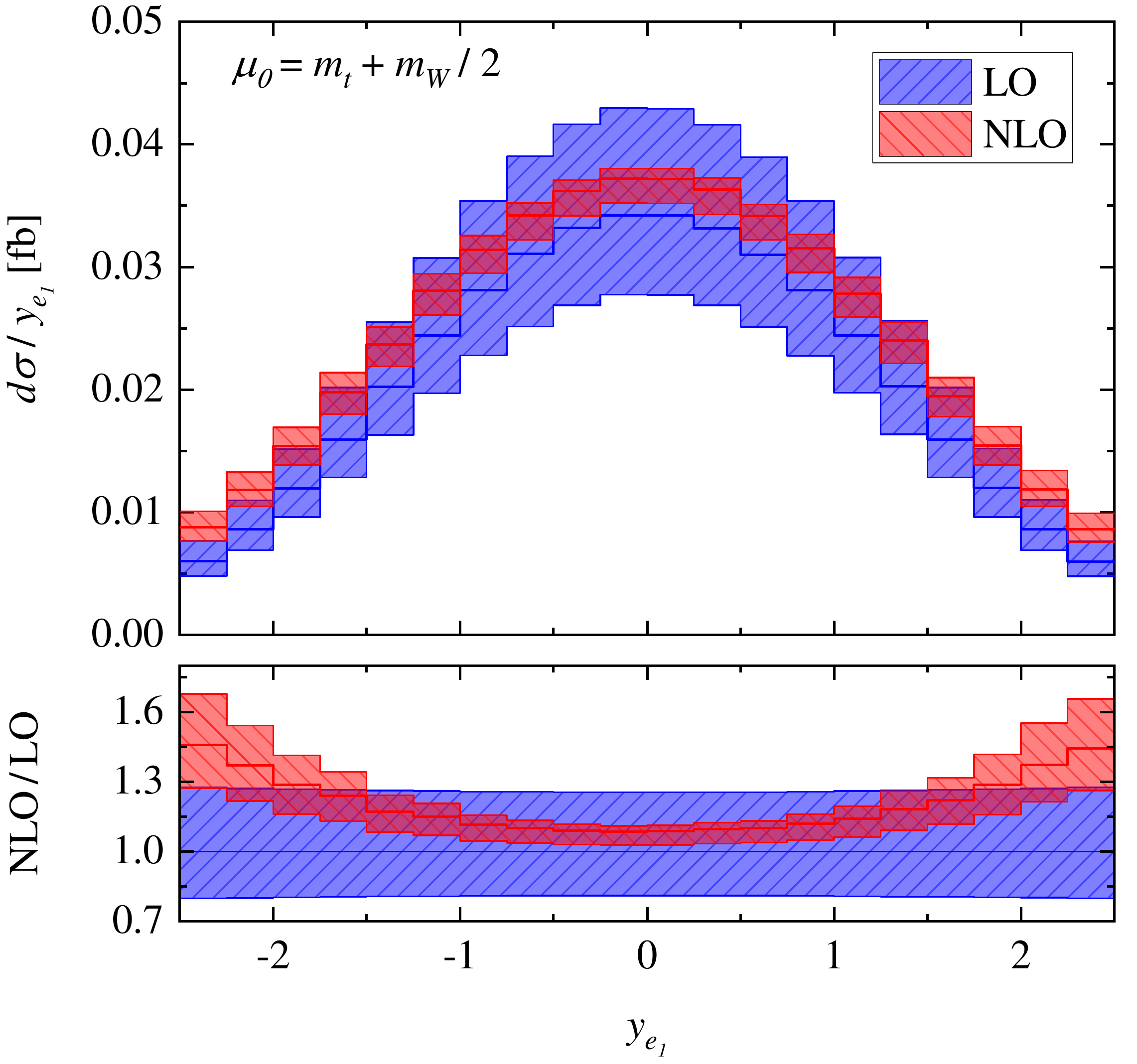}
   \includegraphics[width=0.49\textwidth]{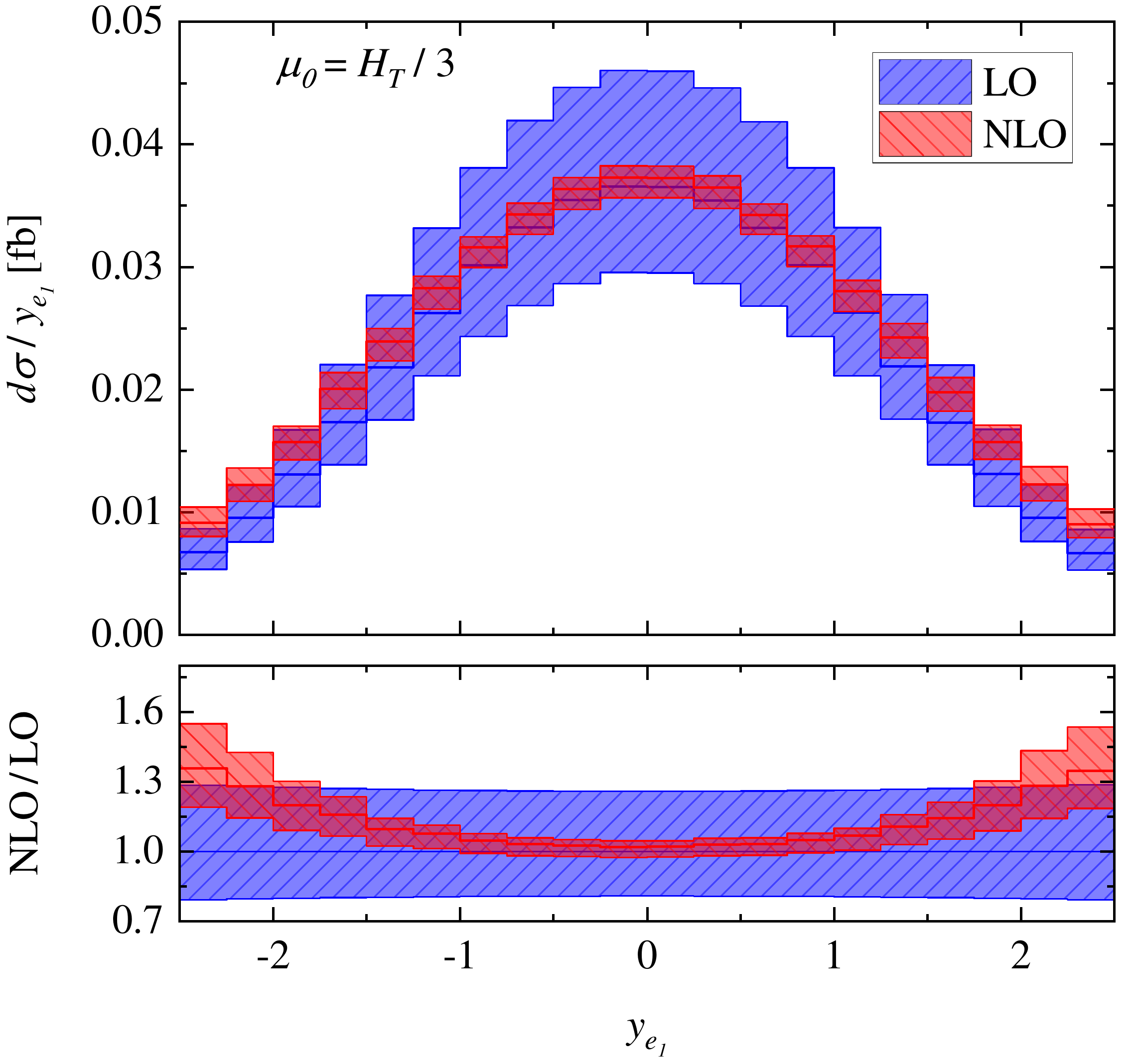}
   \\
 \vspace{0.4cm}
      \includegraphics[width=0.49\textwidth]{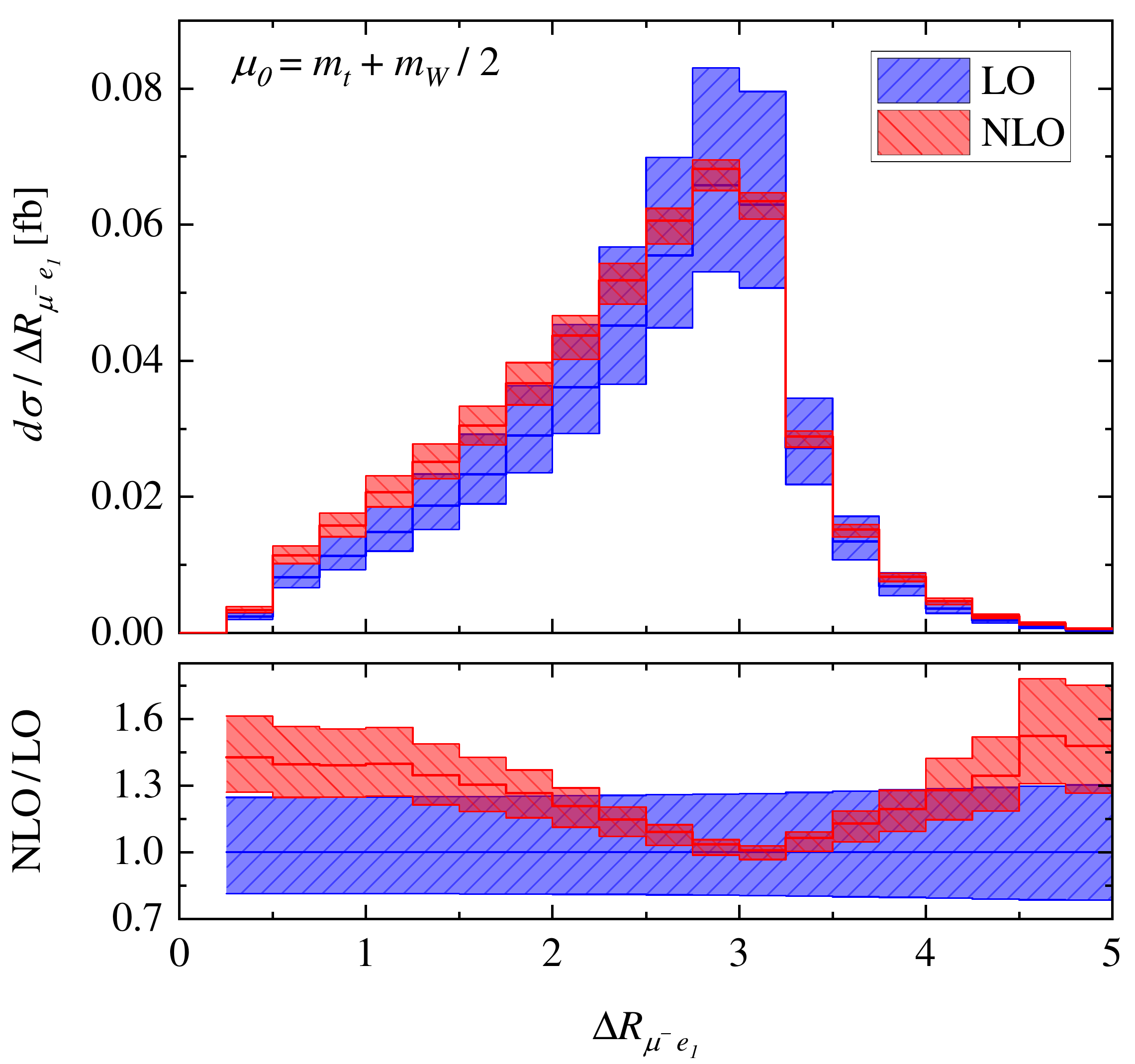}
   \includegraphics[width=0.49\textwidth]{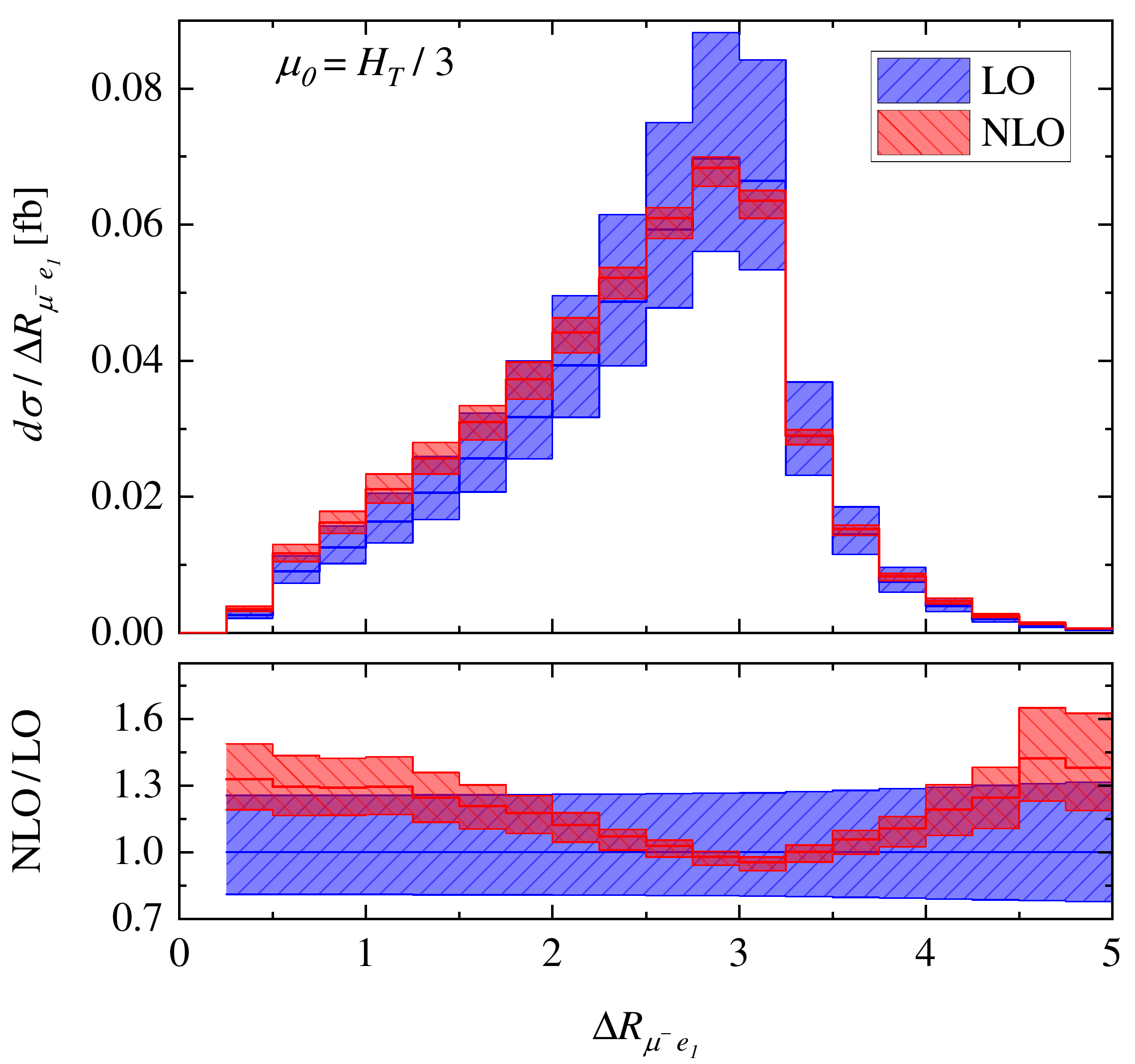} 
\end{center}
\caption{\it   As in Figure \ref{fig:diff1} but for the
    $y_{e_1}$ and $\Delta R_{\mu^- e_1}$ distributions.}
\label{fig:diff3}
\end{figure}
%

In the next step we present the transverse momentum distribution of
the hardest positron, denoted as $p_{T, \, e_1}$, and the invariant
mass of the hardest positron and the muon, labeled as $M_{e_1
\mu^-}$. Both observables are depicted in Figure \ref{fig:diff2}. For
the $p_{T, \, e_1}$ differential cross section distribution with
$\mu_0=m_t+m_W/2$ NLO QCD corrections in the range from $+25\%$ to
$-15\%$ are obtained. Once the kinematic dependent scale choice is
used instead we have rather constant positive corrections of the order
of $10\%$. Also here similar results are observed for the $M_{e_1
\mu^-}$ differential cross section distribution. The resulting
uncertainties for both observables are below $10\%$ independently of
the chosen scale.

Finally, in Figure \ref{fig:diff3} we show  the rapidity of the hardest
positron, $y_{e_1}$, and the separation in the rapidity-azimuthal
angle plane between the muon and the hardest positron, $\Delta
R_{\mu^- e_1}$. Using $\mu_0=m_t+m_W/2$ for $y_{e_1}$ we receive
positive $10\%-20\%$ NLO QCD corrections in the central rapidity
regions. When approaching the forward and backward regions of the
detector these corrections increase rapidly up to even $45\%$. The
situation is once again improved by the dynamical scale choice. In the
central rapidity regions higher-order corrections are only up to
$10\%$ whereas for the forward and backward regions they increase to
$30\%-35\%$. Theoretical uncertainties follow the same pattern. They
are rather small for $|y_{e_1}|<1.5$ of the order of $5\%-10\%$ and
moderate, up to $20\%$, for $|y_{e_1}|\in (1.5 -2.5)$ independently of
the scale choice.  Similar conclusions can be drawn in the case of
$\Delta R_{\mu^- e_1}$. For both observables the differential ${\cal
K}$-factors have large variations.

Overall, the introduction of the dynamical scale stabilises the high
$p_T$ tails of various dimensionful observables and generally provides 
smaller NLO QCD corrections as well as theoretical uncertainties.  We
observe NLO QCD effects up to $10\%-20\%$ and the theoretical
uncertainties due to scale dependence  below $10\%$. For
various dimensionless (angular) cross section distributions the
situation is similar in the central rapidity regions of the
detector. Higher order effects are amplified once the more forward and
backward regions are examined instead.  Independently, in many cases
that we have examined, the differential ${\cal K}$-factors are far
from flat curves, which implies that the NLO QCD corrections have to
be always taken into account to properly model the kinematics of the
process.

%
\subsection{PDF uncertainties}
%

%
\begin{figure}[h!]
\begin{center}
  \includegraphics[width=0.49\textwidth]{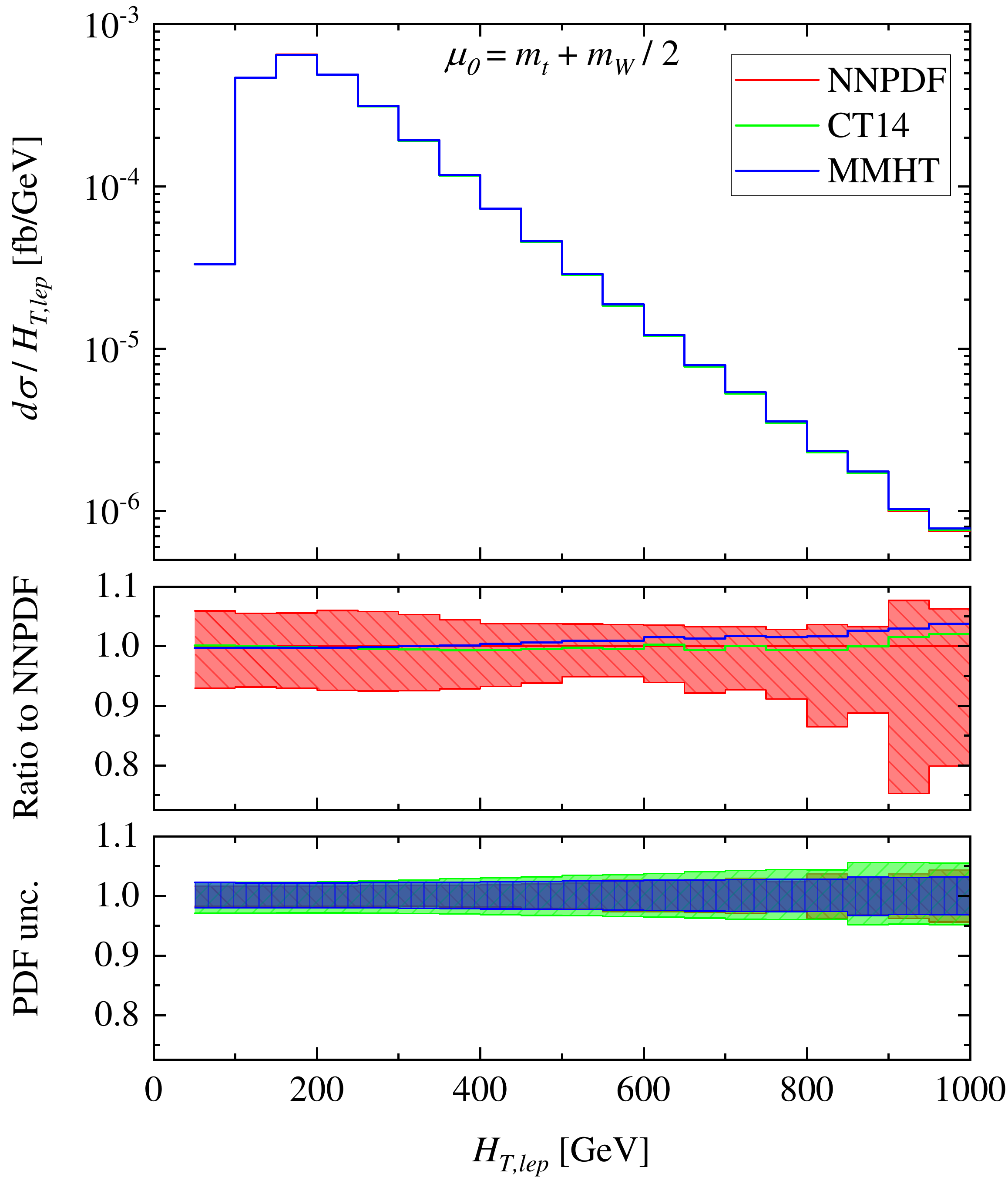}
  \includegraphics[width=0.49\textwidth]{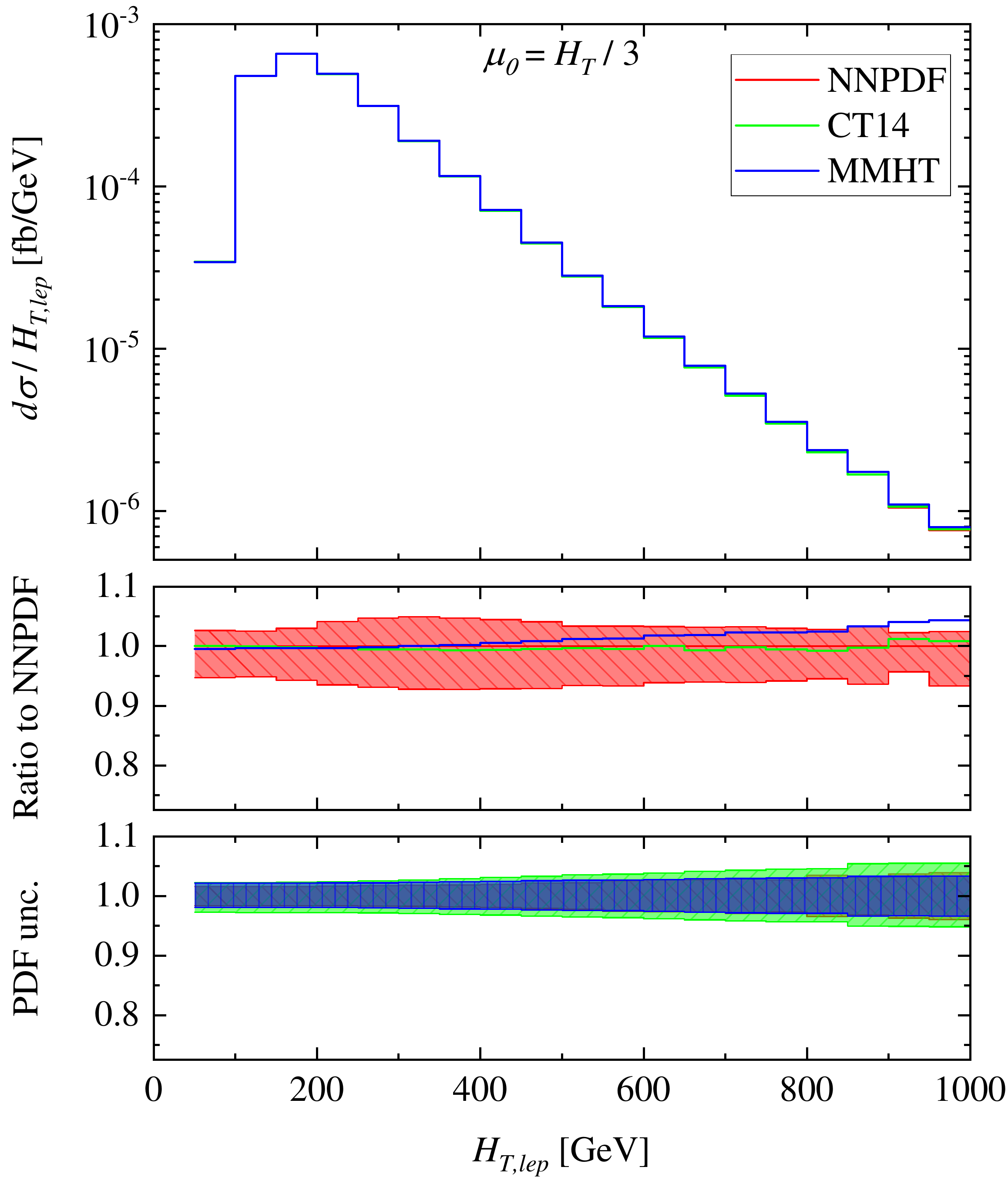}\\
  \vspace{0.4cm}
    \includegraphics[width=0.49\textwidth]{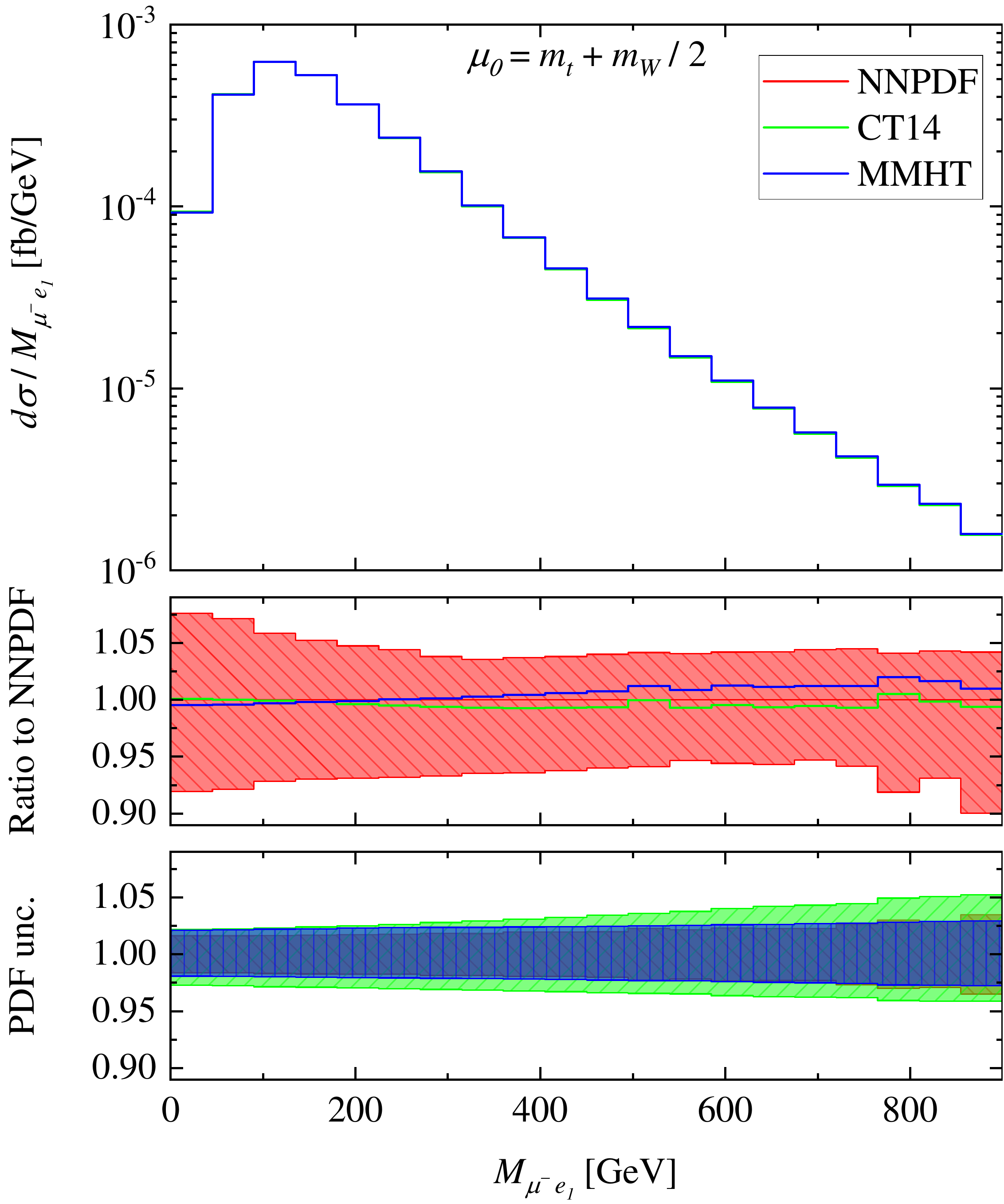}
   \includegraphics[width=0.49\textwidth]{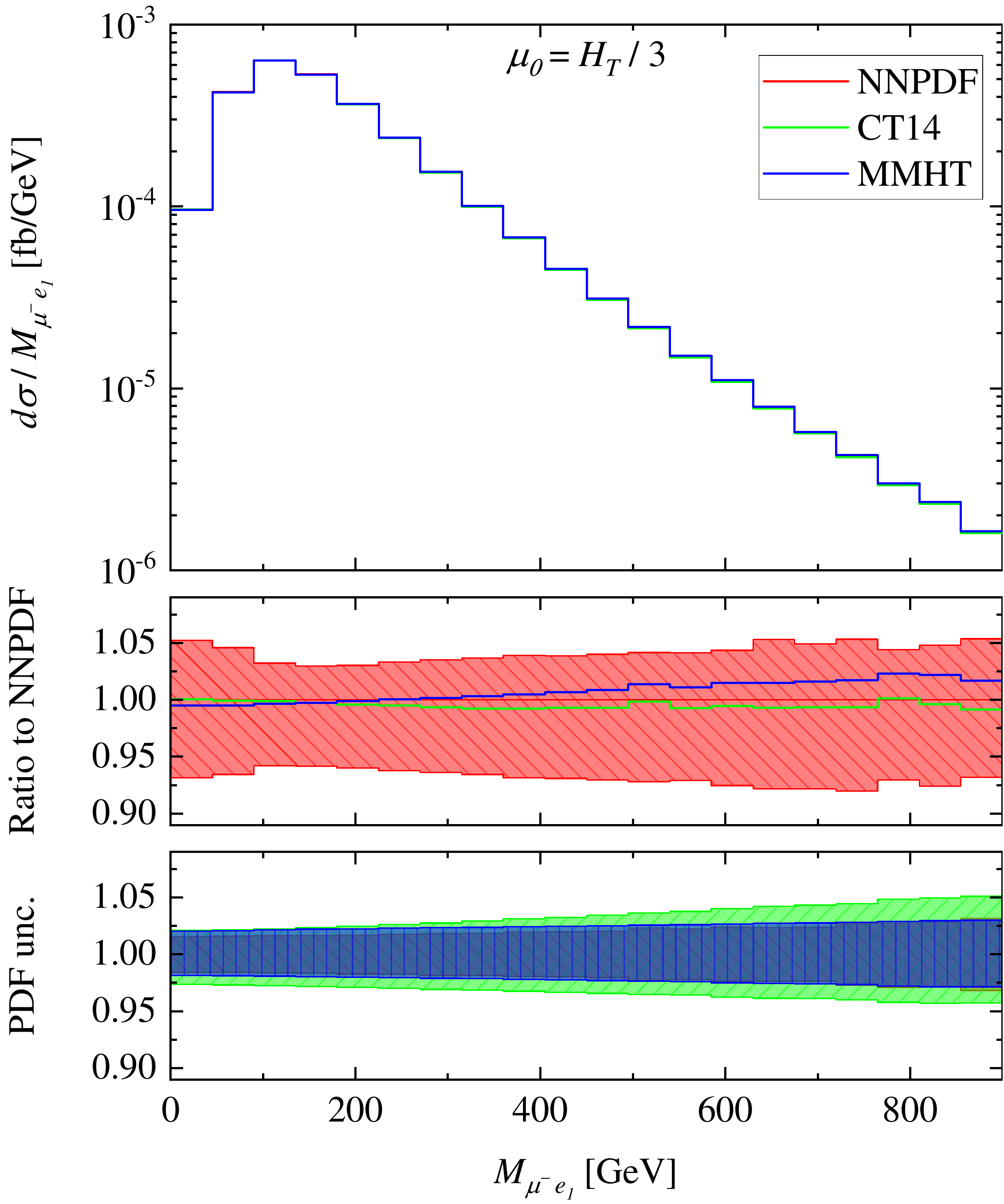}   
\end{center}
\caption{\it
  Differential cross section distributions for the $pp\to e^+\nu_e\,
\mu^-\bar{\nu}_\mu\, e^+\nu_e \, b\bar{b} +X$ process at the LHC with
$\sqrt{s} = 13$ TeV as a function of $H_T^{lep}$ and $M_{\mu^-
e_1}$. The upper plot shows the absolute NLO QCD predictions for three
different PDF sets with $\mu_R=\mu_F=\mu_0$.  The middle panel
displays the ratio to the result with the default NNPDF3.0 PDF set as
well as its scale dependence. The lower panel presents the internal
PDF uncertainties calculated separately for each PDF set.}
\label{fig:pdf1}
\end{figure}
%
%
\begin{figure}[t]
\begin{center}
   \includegraphics[width=0.49\textwidth]{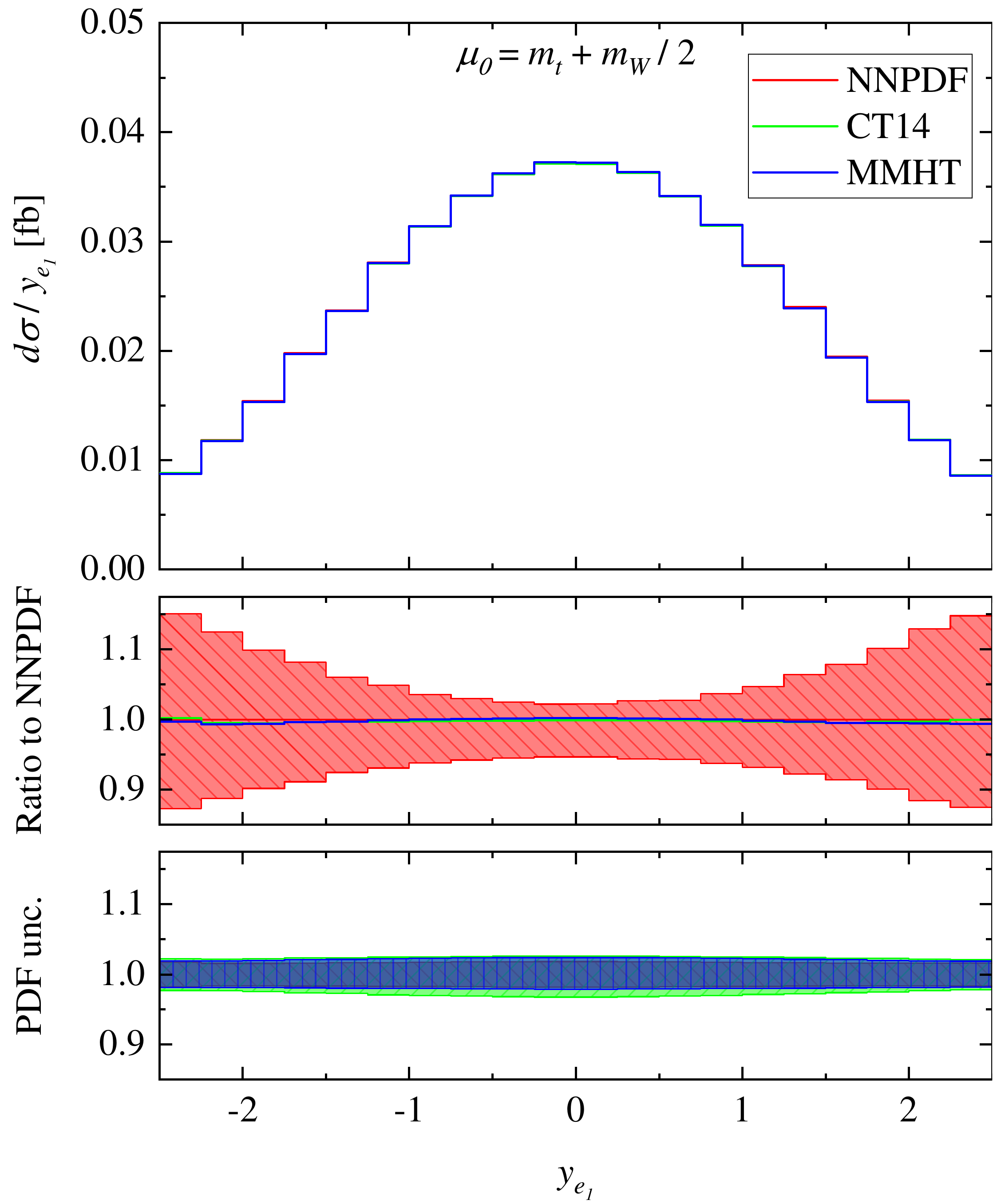}
   \includegraphics[width=0.49\textwidth]{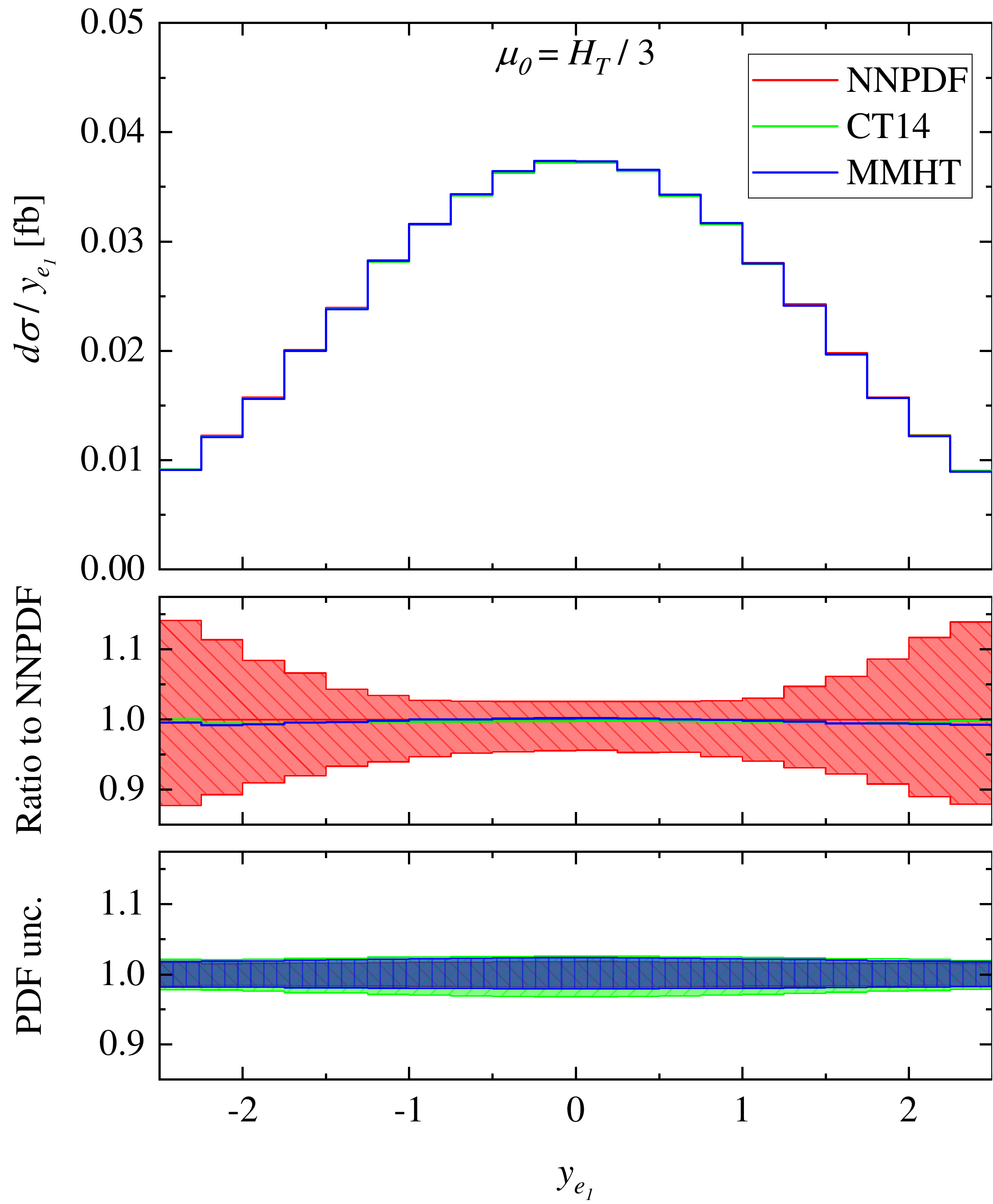}
\end{center}
\caption{\it   As in Figure \ref{fig:pdf1} but for the
    $y_{e_1}$  distribution.}
\label{fig:pdf2}
\end{figure}
%
%

To fully assess the theoretical uncertainties inherent in our
predictions, we shall examine the PDF uncertainties at the
differential level. We have already checked that the latter are below
the uncertainties stemming from scale variation for the integrated
fiducial cross sections. We would like to confirm these findings
differentially for a few observables. We concentrate on three
differential cross section distributions already shown in the previous
section, namely $H_T^{lep}$, $p_{T, \, e_1}$ and $y_{e_1}$. We plot
them afresh for three different PDF sets, CT14, MMHT14 and
NNPDF3.0. Each plot consists of three parts. The upper panel shows the
absolute NLO prediction for three different PDF sets at the central
scale value, $\mu_0$. The middle panel displays the NLO scale
dependence band normalised to the NLO prediction for $\mu_0$ and the
default NNPDF3.0 PDF set. Also shown is the ratio of NLO QCD
predictions generated for the CT14 and MMHT14 PDF set to NNPDF3.0. The
lower panel gives the internal PDF uncertainties for each PDF set
separately, normalised to the corresponding NLO prediction as obtained
with $\mu_R=\mu_F=\mu_0$.

We begin with the differential cross section distribution as a
function of $H_T^{lep}$ shown in Figure \ref{fig:pdf1}. For the fixed
scale choice the PDF uncertainties are of the order of $5\%$, thus, negligible
when contrasted with the theoretical uncertainties from the scale
dependence. Additionally, the differences between results obtained for
various PDF sets are similar in size to the internal PDF
uncertainties. For the dynamical scale choice the PDF uncertainties
and the scale dependence can be of a similar size, especially in the
high $p_T$ regions of the phase space.

For the invariant mass of the muon and the hardest positron, also
given in Figure \ref{fig:pdf1}, the PDF uncertainties are again only
up to $5\%$. Thus, they are smaller than the scale dependence in the
whole plotted range independently of the scale choice.

Finally, the dominance of the scale dependence is even more pronounced
for the rapidity distribution of the hardest positron, presented in
Figure \ref{fig:pdf2}. In this case the PDF uncertainties are well
below $3\%$, therefore, completely negligible when compared with
theoretical uncertainties due to the scale dependence. Moreover, the
differences between various PDF sets at the central scale value,
$\mu_0$, are insignificant. These findings are independent of the
scale choice.

To summarise this part, apart from the high $p_T$ phase space regions
for a few observables the theoretical uncertainties due to the scale
dependence are the dominant source of the theoretical systematics also
for the differential cross sections distributions at NLO in QCD.

%
\subsection{Off-shell versus on-shell top quark decay modelling}
%
%

In this part of the paper we shall examine the size of the
non-factorisable corrections for the $pp\to e^+ \nu_e \, \mu^-
\bar{\nu}_\mu \, e^+ \nu_e \, b\bar{b} +X$ process within our setup.
The non-factorisable corrections vanish in the limit $\Gamma_t/m_t \to
0$, which characterises the NWA. Therefore, to inspect them closely we
compare the NLO QCD results with the complete top-quark off-shell
effects included with the calculations in the NWA.  The latter results
are also generated with the help of the \textsc{Helac-NLO} MC program,
that has recently been extended to provide theoretical predictions in
this approximation \cite{Bevilacqua:2019quz}. The NWA results are
divided in two categories: the full NWA and the ${\rm NWA}_{\rm
LOdecay}$. The full NWA comprises NLO QCD corrections to both the
$t\bar{t}W^\pm$ production and the subsequent top-quark decays
preserving at the same time the $t\bar{t}$ spin correlations. The
${\rm NWA}_{\rm LOdecay}$ case contains the results with NLO QCD
corrections to the production stage only, whereas the top-quark decays
are calculated at LO.  For consistency the NWA result with
the LO top-quark decays is calculated with $\Gamma_{t,\, {\rm
NWA}}^{\rm LO}$. The LO and NLO theoretical predictions for the three
cases are listed in Table \ref{tab:integarted}.  Also provided are the
theoretical uncertainties due to scale dependence. All results are
evaluated for the default NNPDF3.0 PDF sets. To ensure consistency in
the comparison the unexpanded NWA results are used \footnote{ For
consistency in the NWA result the top quark width $\Gamma_t^{\rm
NLO}$, which appears in $\sigma^{\rm NLO}_{t\bar{t}W^\pm}$ as the
factor $(\Gamma_t^{\rm NLO})^{-2}$, should also be computed in series
of $\alpha_s$. In our NLO results in the NWA, however, the top quark
width is not expanded since this procedure can not be directly applied
to the full off-shell calculation.}.  We have checked,
however, that the expanded results are slightly smaller.  The
difference between the expanded and unexpanded NWA results is at $3\%$
level for $\mu_0=H_T/3$ and around $4\%$ for $\mu_0=m_t+m_W/2$.

For the $pp\to e^+\nu_e \, \mu^- \bar{\nu}_\mu \, e^+ \nu_e \, b
\bar{b} +X$ process the complete top-quark off-shell effects change
the  integrated NLO fiducial cross section by less than $0.2\%$
independent of the scale choice. The finding is consistent with the
expected uncertainty of the NWA \cite{Fadin:1993kt}, which is of the
order of ${\cal O}(\Gamma_t/m_t) \approx 0.8\%$ for the inclusive
observables. Having the results in the NWA${}_{\rm LOdecay}$ to our
disposal we can additionally observe that the NLO QCD corrections to
top-quark decays are negative and at the level of $3\%$ for the fixed
scale choice. They increase up to $5\%$ when the dynamical scale is
used instead. Also provided in Table \ref{tab:integarted} are the
theoretical uncertainties due to scale dependence. They are given for
all three cases to help us to investigate whether theoretical
uncertainties are underestimated when various approximations for the
top-quark production and decays are employed instead of the full
description. When comparing the full off-shell case with the full NWA
we notice that theoretical uncertainties are similar, consistently
below $6\%-7\%$ independently of the scale choice.  For the
NWA${}_{\rm LOdecay}$ case, however, they rise up to $10\%-11\%$. We
observe that adding NLO QCD corrections to decays compensates part of
the scale dependence of the cross section with the corrections in the
production.

In summary, both the complete top-quark off-shell effects
and the NLO QCD corrections to top-quark decays are rather small for
the integrated fiducial cross sections. They are consistently within
the NLO theoretical uncertainty estimates for the $pp\to e^+\nu_e \,
\mu^- \bar{\nu}_\mu \, e^+ \nu_e \, b \bar{b} +X$. Additionally, we
note that the full NWA results match better the complete off-shell
predictions on a scale-by-scale basis. Regardless of the
considerations on the scale dependence reduction, the theoretical
description of $t\bar{t}W^\pm$ can only benefit from a more accurate
modelling of of top-quark decays.
%
\begin{table}[t]
  \begin{center}
\begin{tabular}{lcc}
  \hline \hline
  &&\\
  \textsc{Modelling Approach} & $\sigma^{\rm LO}$ [{\rm ab}]
                              & $\sigma^{\rm NLO}$ [{\rm  ab}]
  \\[0.2cm]
  \hline \hline
  &&\\
  full off-shell $(\mu_0=m_t+m_W/2)$ & $106.9^{\, +27.7 \, (26\%)}_{\,
                                       -20.5\, (19\%)}$
                              &  $123.2^{\, +6.3\, (5\%)}_{\, -8.7 \, (7\%)}$\\[0.2cm]
  full off-shell $(\mu_0=H_T/3)$ & $115.1^{\, +30.5\, (26\%)}_{\,
                                   -22.5\, (20\%)}$&
$124.4^{\, +4.3 \, (3\%)}_{\, -7.7 \, (6\%)}$
  \\[0.2cm]
  \hline\hline
 &&\\ 
        NWA  $(\mu_0=m_t+m_W/2)$ & $106.4^{\, +27.5\, (26\%)}_{\,
                                   -20.3\, (19\%)}$ &
       $123.0^{\, +6.3\, (5\%)}_{\,-8.7\, (7\%)}$                         \\[0.2cm]
        NWA  $(\mu_0=H_T/3)$ & $115.1^{\,+30.4\, (26\%)}_{\, -22.4\, (19\%)}$&
                    $124.2^{\,+4.1\, (3\%)}_{\, -7.7\, (6\%)}$           
  \\[0.2cm]
   \hline \hline
  &&\\
  NWA${}_{\rm LOdecay}$ $(\mu_0=m_t+m_W/2)$  &&
                                                $127.0^{\,+14.2\,(11\%)}_{\,-13.3\, (10\%)}$ \\[0.2cm]
        NWA${}_{\rm LOdecay}$ $(\mu_0=H_T/3)$  &&  $130.7^{\,+13.6\,
                                                  (10\%)}_{\,- 13.2\, (10\%)}$\\[0.2cm]
 \hline     \hline                                             
\end{tabular}
\end{center}
\caption{\label{tab:integarted}\it Integrated fiducial cross
sections for the $pp\to e^+\nu_e \, \mu^- \bar{\nu}_\mu e^+ \nu_e \, b
\bar{b} +X$ process at the LHC with $\sqrt{s}=13$ TeV. Results for
various approaches for the modelling of top quark production and decays are
listed. Theoretical uncertainties as obtained from the scale
dependence are also provided. The NNPDF3.0 PDF sets are employed.}
\end{table}
%
%
\begin{figure}[t]
\begin{center}
  \includegraphics[width=0.49\textwidth]{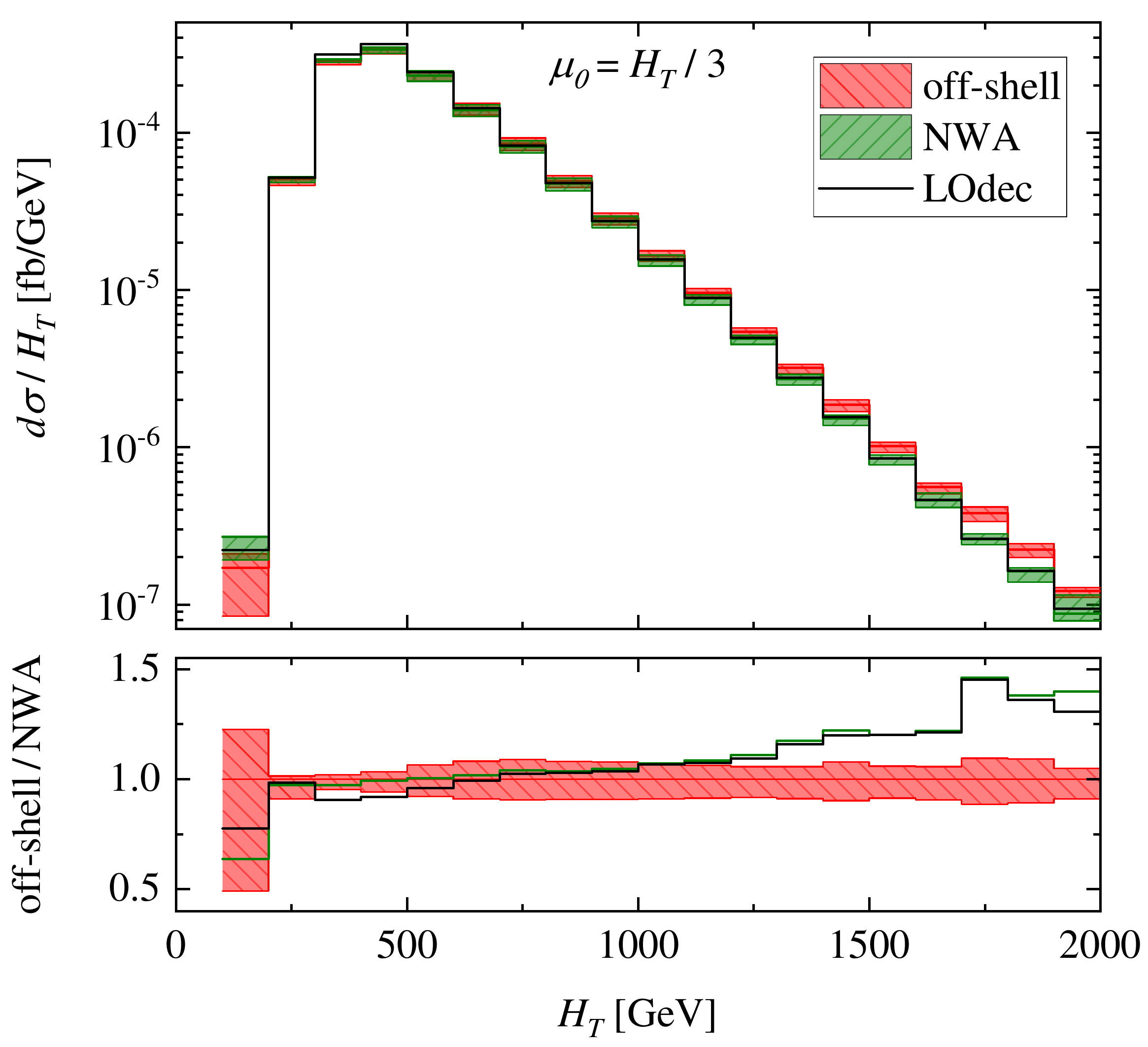}
 \includegraphics[width=0.49\textwidth]{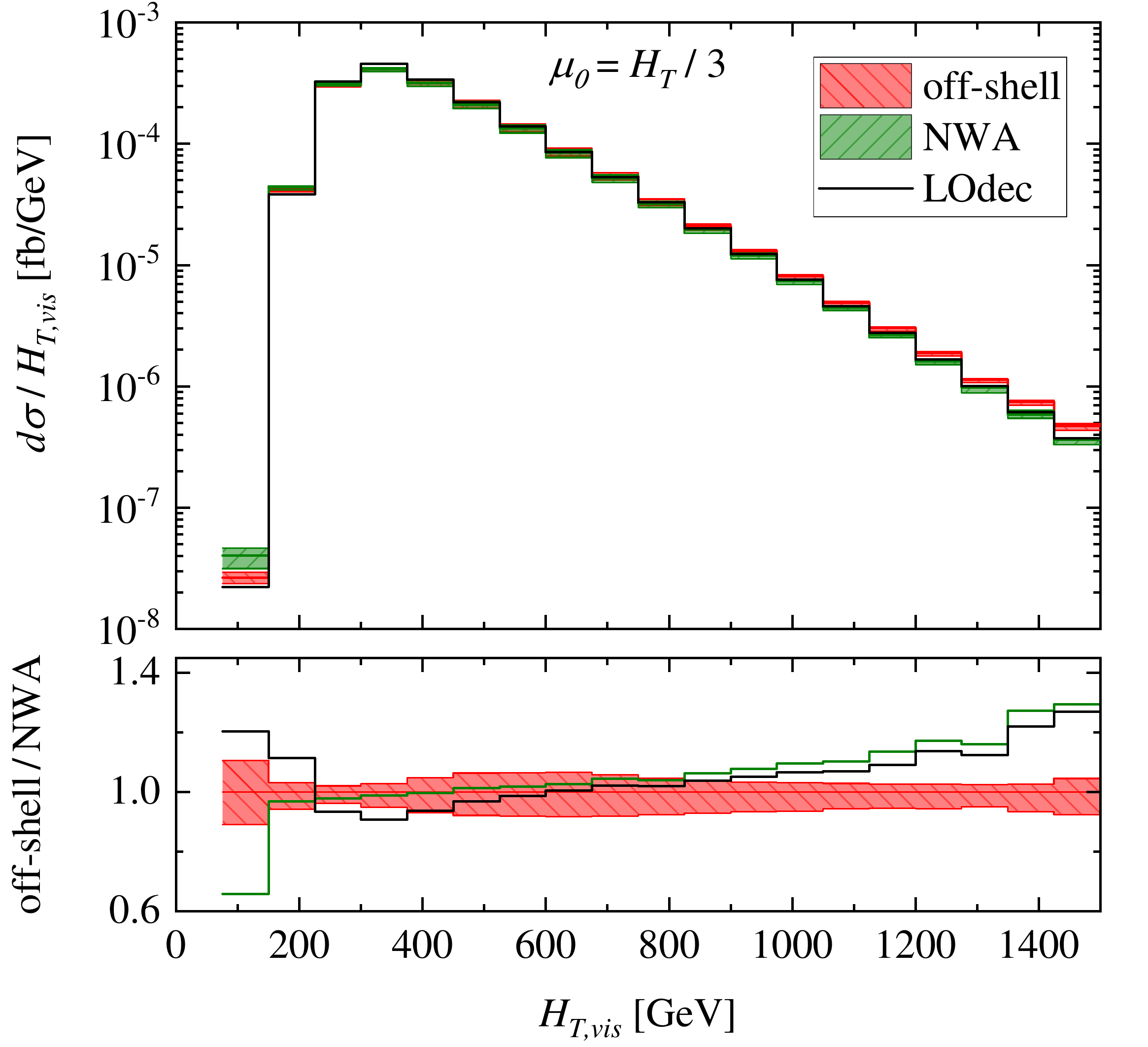}
  \\
  \vspace{0.4cm}
  \includegraphics[width=0.49\textwidth]{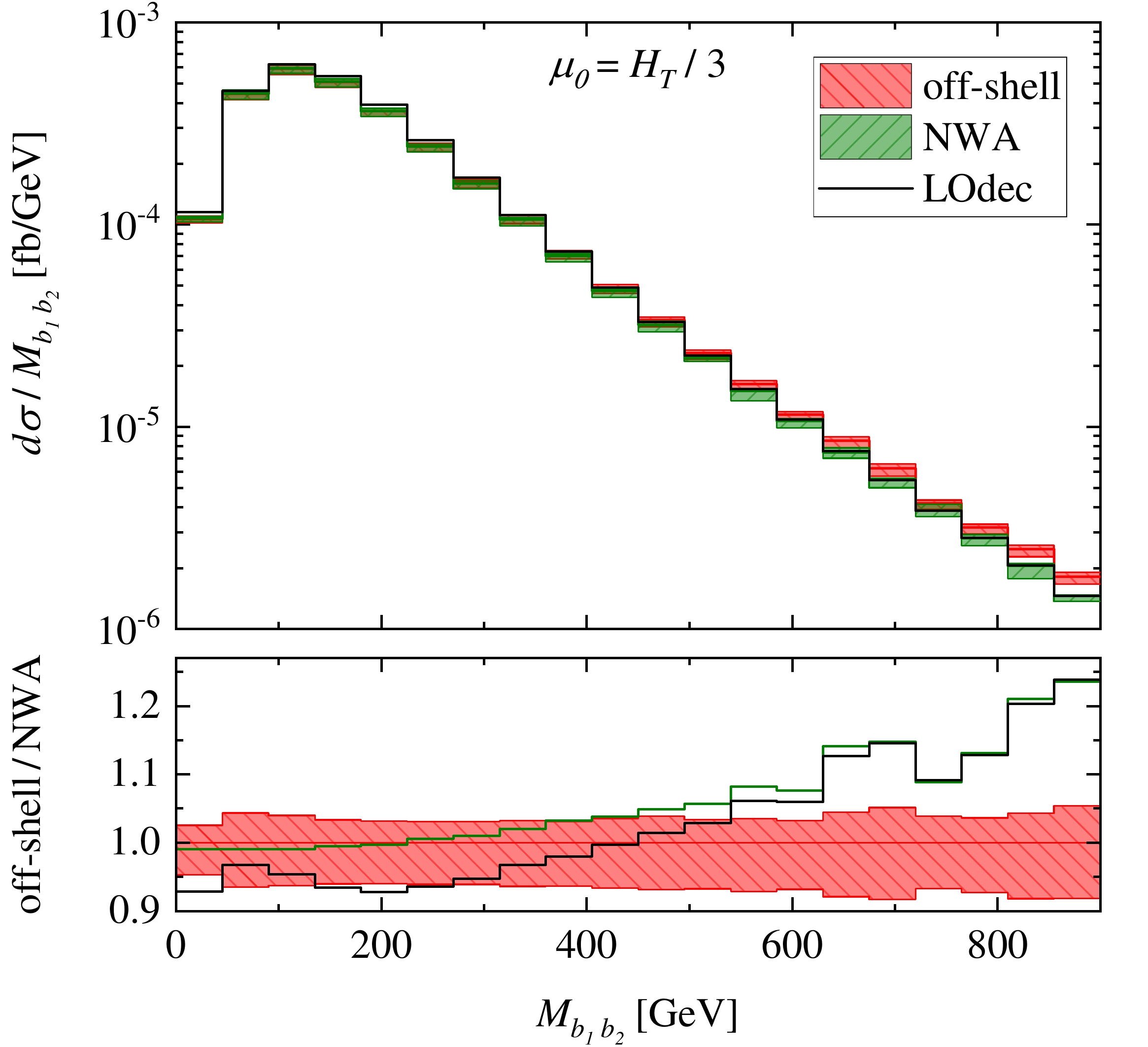}
  \includegraphics[width=0.49\textwidth]{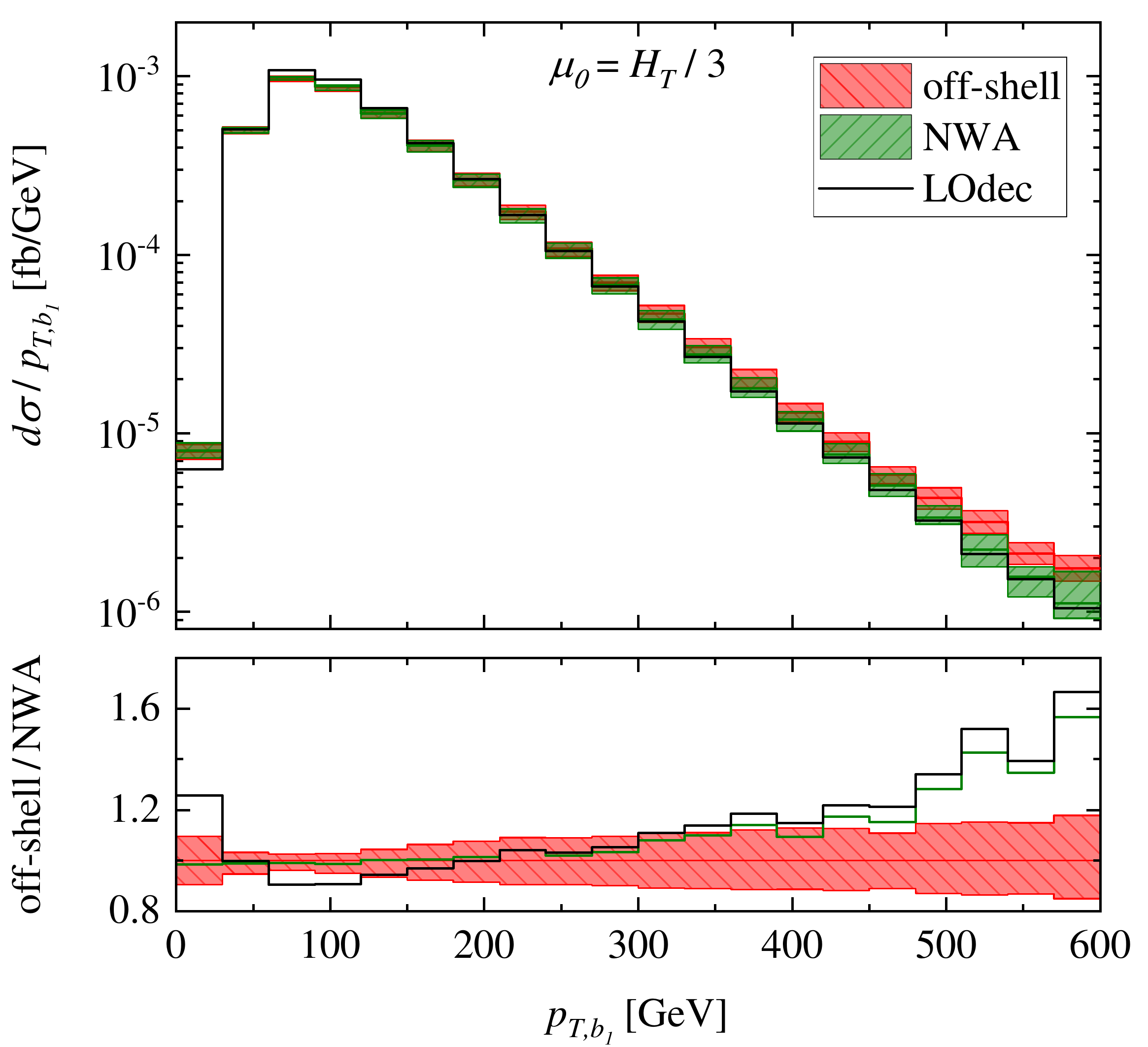}
 \end{center}
\caption{\it  Differential cross section distribution as a
function of $H_T$, $H_T^{vis}$, $M(b_1b_2)$ and $p_T(b_1)$ for the
$pp\to e^+\nu_e \,\mu^- \bar{\nu}_\mu \, e^+\nu_e \,b \bar{b} +X$
production process at the LHC with $\sqrt{s}=13$ TeV.  NLO QCD results
for various approaches for the modelling of top quark production and
decays are shown. We additionally provide theoretical uncertainties as
obtained from the scale dependence for the full off-shell case. Also
plotted are the ratios of the full off-shell result to the two NWA
results. The NNPDF3.0 PDF sets is employed.}
\label{fig:nwa1}
\end{figure}
%
%

A completely different picture emerges when various differential
(fiducial) cross section distributions are analysed at the NLO level
in QCD.  In Figure \ref{fig:nwa1} we exhibit $H_T^{vis}$ and $H_T$.
The latter is defined in Eq.~\eqref{ht}. Also shown in Figure
\ref{fig:nwa1} are the invariant mass of the two $b$-jets, $M_{b_1
b_2}$, and the transverse momentum of the hardest $b$-jet,
$p_{T,\,b_1}$. The same three theoretical descriptions, i.e.  the full
NWA, the NWA${}_{\rm LOdecay}$ and the full off-shell case, are
plotted for the dynamical scale choice and the default NNPDF3.0 PDF
set.  We refrain from presenting differential results for
$\mu_0=m_t+m_W/2 $ because, as we have seen, this scale choice is not
appropriate for differential description of the $pp\to e^+\nu_e \,
\mu^- \bar{\nu}_\mu \, e^+\nu_e \, b \bar{b} +X$ process.  For the
full off-shell case we additionally display the theoretical
uncertainties as obtained from the scale dependence since we are
interested in effects that are outside the NLO uncertainties bands.
The upper plots show the absolute predictions at NLO in QCD, whereas
the bottom plots exhibit the ratios of the full off-shell result to
the two NWA results.

At the tails of the $H_T^{vis}$ distribution we observe that top-quark
off-shell effects increase up to $30\%$. This is well above the
theoretical uncertainties due to scale dependence that for a majority
of dimensionful observables are around $\pm 10\%$. Furthermore, at the
beginning of the spectrum above the kinematical cutoff of $H_T^{vis}
\approx 125$ GeV we can notice large discrepancies between the full
NWA description and the NWA${}_{\rm LOdecay}$ case. They are visible
up to about $400$ GeV, thus, in the regions that are currently
scrutinised by the ATLAS and CMS experiments. Also in this region of
the $H_T^{vis}$ differential cross section distribution the top-quark
off-shell effects are substantial, of the order of $20\%-35\%$.
Similar conclusions can be drawn for $H_T$. In the following we
examine the kinematics of the $b$-jets. For the invariant mass of two
$b$-jets the top-quark off-shell effects are up to $25\%$, whereas in
the case of the transverse momentum of the hardest $b$-jet they are as
large as $60\%-70\%$. For the small values of $M_{b_1 b_2}$ and
$p_{T,\,b_1}$ we can notice $10\%-25\%$ effects.  For the central
value of the scale substantial differences between the full NWA
description and the NWA${}_{\rm LOdecay}$ case are visible also for
these two observables. This highlights  the importance of the proper
modelling of top-quark decays for this process.

In conclusion, in the case of various (dimensionful) differential
cross sections, non-negligible top-quark off-shell effects are present
in various phase-space regions. Substantial differences between the
two versions of the NWA results are additionally observed. Taking into
account that a priori it is not possible to estimate the size of these
effects and which phase space regions are particularly affected a very
careful examination based on the full theoretical description should
be performed on a case-by-case basis. For that reason the complete
top-quark off-shell effects should be included at the differential
level in  future comparisons between theoretical predictions and
experimental data.

%
\section{\boldmath  Phenomenological results for $t\bar{t}W^-$}
\label{sec:ttwm}
%
%

In this section we would like to present the results for the
$t\bar{t}W^-$ process with the complete top quark and $W$ gauge boson
off-shell effects included. As mentioned in the introduction,
however, only theoretical predictions for the integrated fiducial
cross section will be shown. The main reason is not to
extend the manuscript length unnecessarily taking into account that
the NLO QCD effects for $t\bar{t}W^+$ and $t\bar{t}W^-$ are very
similar.

With the input parameters and cuts specified before, we arrive at the
following predictions for the $pp\to e^-\bar{\nu}_e \,\mu^+ \nu_\mu \,
e^-\bar{\nu}_e \,b \bar{b} +X$ process using the default NNPDF3.0 PDF
sets and the fixed scale choice
\begin{equation}
  \begin{split}
&\sigma^{\rm LO}_{e^- \bar{\nu}_e\,
\mu^+ \nu_\mu\, e^- \bar{\nu}_e \, b\bar{b}} \left({\rm NNPDF3.0},
\mu_0=m_t+m_W/2\right)= 57.2^{\,+14.9 \, (26\%)}_{\, -11.0 \, (19\%)}
  \, {\rm [scale]} \, {\rm ab}\,,
\\[0.2cm]
&\sigma^{\rm NLO}_{e^- \bar{\nu}_e\, 
\mu^+ {\nu}_\mu\, e^- \bar{\nu}_e \, b\bar{b}} \left({\rm NNPDF3.0},
\mu_0=m_t+m_W/2 \right)=  68.0^{\, +4.8  \,(7\%)}_{\, -5.5 \, (8\%)} \, {\rm
[scale]} \, {}^{\,+1.2 \, (2\%)}_{\, -1.2\, (2\%)} \, {\rm [PDF]} \, {\rm ab}\,.
\end{split}
\end{equation}  
When the MMHT14 PDF sets are employed instead the following
results are reported 
\begin{equation}
  \begin{split}
&\sigma^{\rm LO}_{e^- \bar{\nu}_e\,
\mu^+ \nu_\mu\, e^- \bar{\nu}_e \, b\bar{b}} \left({\rm MMHT14},
\mu_0=m_t+m_W/2\right)= 55.6^{\,+14.5  \,(26\%)}_{\,-10.7\, (19\%)}
 \, {\rm [scale]} \, {\rm ab}\,,
\\[0.2cm]
&\sigma^{\rm NLO}_{e^- \bar{\nu}_e\, 
\mu^+ {\nu}_\mu\, e^- \bar{\nu}_e \, b\bar{b}} \left({\rm MMHT14},
\mu_0=m_t+m_W/2 \right)= 68.3^{\,+4.6 \, (7\%)}_{\,-5.4 \,(8\%)}  \, {\rm
[scale]} \, {}^{\,+1.6 \, (2\%)}_{\,-1.4 \, (2\%)}\, {\rm [PDF]} \, {\rm ab}\,.
\end{split}
\end{equation}  
Theoretical  predictions for the CT14 PDF sets are given by
\begin{equation}
  \begin{split}
&\sigma^{\rm LO}_{e^- \bar{\nu}_e\,
\mu^+ \nu_\mu\, e^- \bar{\nu}_e \, b\bar{b}} \left({\rm CT14},
\mu_0=m_t+m_W/2\right)= 52.4^{\,+13.4 \, (26\%)}_{\,- 9.9\, (19\%)}
 \, {\rm [scale]} \, {\rm ab}\,,
\\[0.2cm]
&\sigma^{\rm NLO}_{e^- \bar{\nu}_e\, 
\mu^+ {\nu}_\mu\, e^- \bar{\nu}_e \, b\bar{b}} \left({\rm CT14},
\mu_0=m_t+m_W/2 \right)= 66.7^{\,+4.4 \, (7\%)}_{\,-5.3 \, (8\%)}\, {\rm
[scale]} \, {}^{\,+1.7 \, (3\%)}_{\,-2.3 \, (4\%)}\, {\rm [PDF]} \, {\rm ab}\,.
\end{split}
\end{equation}  
The integrated fiducial cross section for $pp \to e^- \bar{\nu}_e\, \mu^+
\nu_\mu\, e^- \bar{\nu}_e \, b\bar{b} +X$ is about a factor of two
smaller than the one for the $pp \to e^+ {\nu}_e\, \mu^-
\bar{\nu}_\mu\, e^+ {\nu}_e \, b\bar{b} +X$ process.  On the other
hand, the behaviour of the QCD higher-order corrections is rather
similar for both processes as one would expect since they are highly
correlated. Specifically, the NLO QCD corrections are positive and
moderate of the order of $19\%$ for the default NNPDF3.0 set. They
increase up to  $23\%\, (27\%)$ for MMHT14 (CT14). The size of
theoretical uncertainties due to scale variation and PDFs is alike.
%
\begin{table}[t!]
  \begin{center}
\begin{tabular}{lcc}
  \hline \hline
  &&\\
  \textsc{Modelling Approach} & $\sigma^{\rm LO}$ [{\rm ab}]
                              & $\sigma^{\rm NLO}$ [{\rm  ab}]
  \\[0.2cm]
  \hline \hline
  &&\\
  full off-shell $(\mu_0=m_t+m_W/2)$ & $57.2^{\,+14.9 \, (26\%)}_{\,
                                       -11.0 \, (19\%)}$
                              &  $68.0^{\, +4.8  \,(7\%)}_{\, -5.5 \,
                                (8\%)}$ \\[0.2cm]
  full off-shell $(\mu_0=H_T/3)$ & $62.4^{\,+16.7 \, (27\%)}_{\,
                                   -12.3\, (20\%)}$ & $68.6^{\, +3.5\, (5\%)}_{\,-4.8\, (7\%)}$ 
  \\[0.2cm]
  \hline\hline
 &&\\ 
        NWA  $(\mu_0=m_t+m_W/2)$ & $57.2^{\,+14.9\, (26\%)}_{\,-11.0\,
                                   (19\%)}$ &    $68.0^{\,+4.9\,
                                              (7\%)}_{\, -5.4 \, (8\%)}$                \\[0.2cm]
        NWA  $(\mu_0=H_T/3)$ & $62.6^{\,+16.7 \, (27\%)}_{\,
                               -12.3\, (20\%)}$  &
                                                    $68.7^{\, +3.5
                                                    \,(5\%)}_{\, -4.8
                                                    \, (7\%)}$
  \\[0.2cm]
   \hline \hline
  &&\\
  NWA${}_{\rm LOdecay}$ $(\mu_0=m_t+m_W/2)$  && $69.8^{\, +8.8\,
                                                (13\%)}_{\, -7.8 \, (11\%)}$
                                               \\[0.2cm]
        NWA${}_{\rm LOdecay}$ $(\mu_0=H_T/3)$  && $72.0^{\, +8.3 \,
                                                  (11\%)}_{\,-7.7\, (11\%)}$
  \\[0.2cm]
 \hline     \hline                                             
\end{tabular}
\end{center}
\caption{\label{tab:integarted2}\it Integrated fiducial  cross
sections for the $pp\to e^-\bar{\nu}_e \, \mu^+ \nu_\mu \,
e^-\bar{\nu}_e \, b \bar{b} +X$ process at the LHC with $\sqrt{s}=13$
TeV. Results for various approaches for the modelling of top quark
production and decays are listed. Theoretical uncertainties as
obtained from the scale dependence are also provided. The NNPDF3.0 PDF
sets are employed.}
\end{table}
%
Also in this case the stability test with respect to the $p_T(j_b)$
cut has been performed for the integrated fiducial cross section yielding
excellent theoretical control over higher-order QCD corrections for
this process. For completeness we report on the results for the
dynamical scale choice, $\mu_0=H_T/3$, with $H_T$ given this time by
\begin{equation}
H_T= p_T(\ell_1) + p_T(\ell_2) +p_T(\mu^+) + p_T^{miss} + p_T(j_{b_1})
+ p_T(j_{b_2}) \,,
\end{equation}  
where $\ell_{1,2}=e^-_{1,2}\,$ are the hardest and the softest
electron. For the NNPDF3.0 PDF sets we have
\begin{equation}
  \begin{split}
&\sigma^{\rm LO}_{e^- \bar{\nu}_e\,
\mu^+ \nu_\mu\, e^- \bar{\nu}_e \, b\bar{b}} \left({\rm NNPDF3.0},
\mu_0=H_T/3\right)= 62.4^{\,+16.7 \, (27\%)}_{\, -12.3\, (20\%)}
 \, {\rm [scale]} \, {\rm ab}\,,
\\[0.2cm]
&\sigma^{\rm NLO}_{e^- \bar{\nu}_e\, 
\mu^+ {\nu}_\mu\, e^- \bar{\nu}_e \, b\bar{b}} \left({\rm NNPDF3.0},
\mu_0=H_T/3 \right)= 68.6^{\, +3.5\, (5\%)}_{\,-4.8\, (7\%)}  \, {\rm
[scale]} \, {}^{\,+1.2 \, (2\%)}_{\,-1.2\,  (2\%)}  \, {\rm [PDF]} \, {\rm ab}\,.
\end{split}
\end{equation}  
For the MMHT14 PDF sets  the results are as follows 
\begin{equation}
  \begin{split}
&\sigma^{\rm LO}_{e^- \bar{\nu}_e\,
\mu^+ \nu_\mu\, e^- \bar{\nu}_e \, b\bar{b}} \left({\rm MMHT14},
\mu_0=H_T/3\right)= 60.5^{\,+16.1 \, (27\%)}_{\,-11.9\, (20\%)}
 \, {\rm [scale]} \, {\rm ab}\,,
\\[0.2cm]
&\sigma^{\rm NLO}_{e^- \bar{\nu}_e\, 
\mu^+ {\nu}_\mu\, e^- \bar{\nu}_e \, b\bar{b}} \left({\rm MMHT14},
\mu_0=H_T/3 \right)= 68.9^{\, +3.3 \, (5\%)}_{\,-4.7 \, (7\%)}  \, {\rm
[scale]} \,  {}^{\,+1.6\, (2\%)}_{\, -1.4\, (2\%)} \, {\rm [PDF]} \, {\rm ab}\,.
\end{split}
\end{equation}  
With the CT14 PDF sets the results read
\begin{equation}
  \begin{split}
&\sigma^{\rm LO}_{e^- \bar{\nu}_e\,
\mu^+ \nu_\mu\, e^- \bar{\nu}_e \, b\bar{b}} \left({\rm CT14},
\mu_0=H_T/3 \right)=57.0^{\, +14.9 \,  (26\%)}_{\, -11.0 \, (19\%)}
 \, {\rm [scale]} \, {\rm ab}\,,
\\[0.2cm]
&\sigma^{\rm NLO}_{e^- \bar{\nu}_e\, 
\mu^+ {\nu}_\mu\, e^- \bar{\nu}_e \, b\bar{b}} \left({\rm CT14},
\mu_0= H_T/3 \right)= 67.3^{\, +3.1 \, (5\%)}_{\,-4.6 \, (7\%)}  \, {\rm
[scale]} \, {}^{\, +1.7 \, (3\%)}_{\,-2.3 \, (3\%)}  \, {\rm [PDF]} \, {\rm ab}\,.
\end{split}
\end{equation}  
Finally, in Table \ref{tab:integarted2} we present the integrated
fiducial cross sections for the full off-shell case, the full NWA and
for NWA${}_{\rm LOdecay}$. Theoretical uncertainties as obtained from
scale variations are also provided.  All LO and NLO results are
presented for the default NNPDF3.0 PDF sets. Our findings are much the
same as in the case of the $pp\to e^+ {\nu}_e \, \mu^- \bar{\nu}_\mu \,
e^+ {\nu}_e \, b \bar{b} +X$ production process.

%
\section{Summary and Outlook}
\label{sec:sum}
%
%

In this paper we have calculated NLO QCD corrections to the $e^+ \nu_e
\,\mu^-\bar{\nu}_\mu \, e^+ \nu_e \, b\bar{b}$ and $e^- \bar{\nu}_e \,
\mu^+ {\nu}_\mu \, e^- \bar{\nu}_e \, b\bar{b}$ final states in
$t\bar{t}W^\pm$ production. In the
computation off-shell top quarks have been described by the
Breit-Wigner distribution, furthermore double-, single- as well as
non-resonant top quark contributions along with all interference
effects have been consistently incorporated already at the matrix
element level. We presented our results for the LHC Run II centre of
mass system energy of $\sqrt{s}=13$ TeV for the two scale choices
$\mu_0=m_t+m_W/2$ and $\mu_0=H_T/3$ and the following three PDF sets
NNPDF3.0, MMHT14 and CT14. For the default NNPDF3.0 PDF set with
$\mu_0=m_t+m_W/2$ moderate NLO QCD corrections of the order of $15\%
\, (19\%)$ have been found for the $t\bar{t}W^+$ $(t\bar{t}W^-)$
integrated fiducial cross section. When $\mu_0=H_T/3$ has been
employed instead they are reduced down to $8\% \, (10\%)$
respectively. Detailed studies of the scale dependence of our NLO
predictions have indicated that the residual theoretical uncertainties
due to missing higher-order corrections are below $6\%-8\%$
independently of the scale choice. The PDF uncertainties are up to
$2\%-4\%$ only. Thus, the theoretical uncertainties due to the scale
dependence are the dominant source of the theoretical systematics.

For differential cross section distributions large shape distortions
have been observed in the presence of higher-order QCD effects.  The
non-flat differential ${\cal K}$-factors underlined the importance of
NLO QCD corrections for proper modelling of the process
kinematics. Furthermore, we observed that $\mu_0=m_t+m_W/2$ led to
perturbative instabilities in the TeV regions of various dimensionful
observables.   The introduction of the dynamical scale stabilised the
high $p_T$ tails and generally provided smaller NLO QCD corrections as
well as theoretical uncertainties. For $\mu_0=H_T/3$ we obtained NLO
QCD effects up to $10\% - 20\%$ and the theoretical uncertainties due
to scale dependence are below $10\%$. The latter are the dominant
source of the theoretical systematics.

In addition, the size of the complete top-quark off-shell effects has
been examined. For the integrated fiducial cross sections negligible
effects, that are consistent with the expected uncertainty of the NWA,
have been found. At the differential level, however, large
non-factorisable corrections even up to $60\%-70\%$ have been observed.

Last but not least, the size of NLO QCD corrections to the top-quark
decays has been studied. These corrections were rather small up to
$5\%$ only for the integrated fiducial cross sections. For various differential
distributions, on the other hand, the differences between the full NWA
and the NWA${}_{\rm LOdecay}$ case were substantial especially in the
low $p_T$ regions. The latter phase space regions are currently
scrutinised by the ATLAS and CMS experimental collaborations at the
LHC. Furthermore, for the integrated fiducial cross section we noticed
that the theoretical uncertainties due to scale dependence were alike
for the full off-shell and full NWA case.  They were systematically
below $6\% - 8\%$ showing that the full NWA predictions would not
underestimate or overestimate the theoretical uncertainties as long as
NLO QCD corrections were consistently incorporated at every stage of
the process. Having rather small uncertainties for the $t\bar{t}W^\pm$
process force us to look for other effects, that might be of
comparable size. The latter, comprise for example formally sub-leading
electroweak corrections, which include $tW \to tW$ scattering
\cite{Frederix:2020jzp}. As shown in Ref. \cite{Frederix:2020jzp} the
combined effect of spin correlations in the top-quark pair and
sub-leading electroweak contributions, which were larger than the
so-called NLO electroweak corrections, would enhance the normalisation
of the $t\bar{t}W$ process by approximately $10\%$.

Finally, in the case of NWA${}_{\rm LOdecay}$, i.e. in
the presence of LO top-quark decays, theoretical uncertainties at NLO
in QCD increased to $11\% - 13\%$.   Regardless of the
considerations on the scale dependence reduction, the theoretical
description of $t\bar{t}W^\pm$ can only benefit from a more accurate
modelling of of top-quark decays.

To recapitulate, the non-factorisable NLO QCD corrections
as well as higher-order QCD effects in top-quark decays impacted
significantly the $t\bar{t}W^\pm$ cross section in various phase space
regions.  For these reasons they should both be included in the future
comparisons between theoretical predictions and experimental data.  In
addition, in view of the importance of the $t\bar{t}W^\pm$ process as
background to Higgs boson production in association with the top quark
pairs, more detailed and combined phenomenological studies for
$t\bar{t}W^+$ and $t\bar{t}W^-$ in the multi-lepton channel are a
necessity. We postpone such work for the future.

\acknowledgments{ The research of G.B. was supported by grant K 125105
of the National Research, Development and Innovation Office in
Hungary.

The work of H.B.H. has received funding from the European Research
Council (ERC) under the European Union's Horizon 2020 research and
innovation programme (grant agreement No 772099).

The work of H.B. and M.W.  was supported by the Deutsche
Forschungsgemeinschaft (DFG) under grant 396021762 $-$ TRR 257.

Support by a grant of the Bundesministerium f\"ur Bildung und
Forschung (BMBF) is additionally acknowledged.

Simulations were performed with computing resources granted by RWTH
Aachen University under project {\tt rwth0414.}}

\end{document}